\documentclass[aps,pra,twocolumn,floatfix,superscriptaddress]{revtex4}

\usepackage{amssymb,amsmath,amstext}                
\usepackage{graphicx}                                               
\usepackage{epstopdf}                                               
\usepackage{color}                                                     
\usepackage{bm}                                                        
\usepackage{appendix}                                              
\usepackage[utf8]{inputenc}
\usepackage{bbold}
\usepackage{bbm}
\usepackage{ulem}
\usepackage{latexsym}
\usepackage[colorlinks=true,citecolor=blue,linkcolor=magenta]{hyperref}

\newcommand{\obar}[1]{\mkern 1.5mu\overline{\mkern-1.5mu#1\mkern-1.5mu}\mkern 1.5mu}

\def\be{\begin{equation}}
\def\ee{\end{equation}}
\def\bea{\begin{eqnarray}}
\def\eea{\end{eqnarray}}
\def\ra{\rangle}
\def\la{\langle}
\def\bi{\begin{itemize}}
\def\ei{\end{itemize}}
\def\ben{\begin{enumerate}}
\def\een{\end{enumerate}}

\begin{document} 

\title{Fractional Time Crystals}

\author{Pawe\l{} Matus} 
\affiliation{
Instytut Fizyki imienia Mariana Smoluchowskiego, 
Uniwersytet Jagiello\'nski, ulica Profesora Stanis\l{}awa \L{}ojasiewicza 11, PL-30-348 Krak\'ow, Poland}
\author{Krzysztof Sacha} 
\affiliation{
Instytut Fizyki imienia Mariana Smoluchowskiego, 
Uniwersytet Jagiello\'nski, ulica Profesora Stanis\l{}awa \L{}ojasiewicza 11, PL-30-348 Krak\'ow, Poland}
\affiliation{Mark Kac Complex Systems Research Center, Uniwersytet Jagiello\'nski, ulica Profesora Stanis\l{}awa \L{}ojasiewicza 11, PL-30-348 Krak\'ow, Poland
}

\begin{abstract}
Time crystals are quantum systems which are able to reveal condensed matter behavior in the time domain. It is known that crystalization in time can be observed in a periodically driven many-body system when interactions between particles force a system to evolve with a period which is an integer multiple of a driving period. This phenomenon is dubbed discrete time crystal formation. Here, we consider ultra-cold atoms bouncing on an oscillating atom mirror and show that the system can spontaneously form a discrete time crystal where the ratio of a period of its motion and a driving period is a rational number. This kind of discrete time crystals requires higher order resonant driving which is analyzed here with the help of an original approach. 
\end{abstract}
\date{\today}

\maketitle

\section{Introduction}

It has been known for years that atoms can self-organize and form periodic structures in space. Formation of such space crystals is related to spontaneous breaking of continuous space translation symmetry of the Hamiltonian \cite{Strocchi2005}. Interactions between particles in a solid state system depend on relative distances between them and if we translate all particles by the same arbitrary vector, the Hamiltonian does not change. Consequently, the eigenstates should follow the symmetry requirement and no crystalline structure can be observed if we calculate a single particle probability density for a system prepared in an eigenstate. However, the eigenstates can be fragile to any perturbation and the continuous space translation symmetry can be extremely easily broken and a space crystal emerges. 

In 2012 Frank Wilczek asked the question if a similar phenomenon could be observed in the time domain and the time crystal era began \cite{Wilczek2012,Sacha2017rev}. The original Wilczek's model was not feasible because he assumed a time-independent quantum many-body system in the ground state \cite{Wilczek2012,Bruno2013b,Watanabe2015,Syrwid2017}  
\footnote{A recent paper presents spontaneous breaking of continuous time translation symmetry in a system ground state, see  \cite{Ohberg2018}.}
(for classical time crystals see \cite{Shapere2012,Ghosh2014,Yao2018,Das2018,Alvarez2017,Aviles2017,Flicker2018}). However, soon after that, it was shown that periodically driven quantum many-body systems can reveal spontaneous crystalization of periodic motion \cite{Sacha2015,Khemani16,ElseFTC,Yao2017,Zhang2017,Choi2017}. Periodically driven systems can be prepared in a Floquet eigenstate which evolves with a driving period \cite{Shirley1965}. If interactions between particles of a driven system are sufficiently strong, some of Floquet eigenstates can become Schr\"odinger cat-like states. Anything can destroy a Schr\"odinger cat state and a new state emerges which breaks the time translation symmetry of the Hamiltonian because it evolves with a period which is an integer multiple of the driving period \cite{Sacha2015}. Such spontaneous self-reorganization of motion of a many-body system is dubbed a discrete time crystal because discrete time translation symmetry of the time-periodic Hamiltonian is spontaneously broken \cite{ElseFTC}. Recently it was shown that when discrete time translation symmetry is spontaneously broken, a quantum many-body system can also reveal a time quasi-crystal behavior \cite{Giergiel2018c}.

Research on time crystals is developing rapidly. New theoretical ideas and experimental results are being published and condensed matter physics has migrated to the time domain \cite{sacha15a,delande17,Giergiel2017,Lazarides2017,Russomanno2017,Zeng2017,Nakatsugawa2017,Ho2017,
Huang2017,Gong2017,Wang2017,Iemini2017,Mierzejewski2017,Prokofev2017,Pal2018,Rovny2018,
Rovny2018a,
Giergiel2018,Bomantara2018,Kosior2018,Mizuta2018,Giergiel2018a,Giergiel2018b,Autti2018,
Tucker2018,
Kosior2018a,Yu2018,Sullivan2018,Smits2018,Gambetta2018,Liao2018,Hayata2018,
Surace2018,Kreil2018} (for phase space crystals see \cite{Guo2013,Guo2016,Guo2016a,Liang2017}). Anderson or many-body localization and Mott-insulator phase can be also observed in time \cite{sacha15a,delande17,Giergiel2017,Mierzejewski2017,Giergiel2018a}, topological time crystals \cite{Lustig2018,Giergiel2018b}, time quasi-crystals \cite{Li2012,Huang2017a,Giergiel2018,Autti2018,Giergiel2018c} and exotic condensed matter-like systems in the time domain \cite{Giergiel2018} can be realized.

In the present paper we concentrate on ultra-cold atoms bouncing on an oscillating atom mirror and show that the system is able to form spontaneously a periodic structure in time where the ratio of a period of its motion and a driving period is a rational number. In order to describe such phenomenon higher order resonant dynamics has to be analyzed. The analysis  is carried out with the help of an original approach which is compared with the standard second order perturbation theory \cite{Lichtenberg1992}. The classical analysis is the basis for the quantum many-body approach.

The paper is organized as follows. In Sec.~\ref{sec1} we introduce a single-particle version of the system and perform an analysis of the second order resonances. In Sec.~\ref{sec2} we consider the many-body problem and identify a range of parameters for which the system reveals spontaneous formation of a discrete time crystal. We summarize the results in Sec.~\ref{seccon}.

\section{Single-particle system}
\label{sec1}

We begin with a description of an atom bouncing on an oscillating atom mirror in the presence of the gravitational force \cite{Buchleitner2002,Steane95} (for stationary mirror experiments see \cite{Roach1995,Sidorov1996,Westbrook1998,Lau1999,Bongs1999,Sidorov2002,Fiutowski2013,Kawalec2014}). We concentrate on a classical analysis which allows us to obtain an effective Hamiltonian of a resonantly driven atom. The effective Hamiltonian is then used as the basis for a quantum description.

\subsection{Single-particle Hamiltonian}

The Hamiltonian of an atom bouncing on an oscillating atom mirror, in the one-dimensional model and in the gravitational units \cite{Giergiel2018a}, reads
\be
H=\frac{p^2}{2}+z+F\left(z+\frac{\lambda}{\omega^2}\cos\omega t\right),
\label{h_lab}
\ee
where $F(z)$ describes the mirror, i.e., the profile of the reflecting potential, which oscillates harmonically with the amplitude $\lambda/\omega^2$ and frequency $\omega$. Description of the system is more convenient if we switch from the laboratory frame to the frame that oscillates with the mirror. Then, the mirror does not move but the gravitational acceleration oscillates in time \cite{Giergiel2018a},
\be
H=\frac{p^2}{2}+z+\lambda z\cos\omega t+F\left(z\right).
\label{h}
\ee
We assume that the mirror can be modeled by a hard wall potential located at $z=0$ in the oscillating frame, and therefore we may drop the $F(z)$ in Eq.~(\ref{h}) keeping in mind that motion of an atom is restricted to $z\ge 0$.

Analysis of resonant dynamics of a particle is convenient when we perform a canonical transformation from the Cartesian variables ($p$ and $z$) to the so-called action angle variables ($I$ and $\theta$) of the unperturbed problem \cite{Lichtenberg1992}. In these new canonically conjugate variables, the unperturbed Hamiltonian depends on the action (new momentum) only \cite{Buchleitner2002},
\be
H_0=\frac{p^2}{2}+z=\frac{(3\pi I)^{2/3}}{2},
\label{Iformula}
\ee
and it is straightforward to get a solution of the unperturbed problem. Indeed, the action is a constant of motion, $I=$const., and the angle (which describes positon of a particle on a periodic trajectory) evolves linearly in time, $\theta(t)=\Omega(I)t+\theta(0)$ where 
\be
\Omega(I)=\frac{dH_0(I)}{dI},
\label{Om}
\ee
is the frequency of a periodic motion of a particle. The entire Hamiltonian (\ref{h}) in the action-angle variables reads
\be
H=H_0(I)+\lambda \cos\omega t\sum_nh_n(I)e^{in\theta},
\label{h_action}
\ee
where $h_0(I)=\left(\frac{\pi I}{\sqrt{3}}\right)^{2/3}$ and $h_n(I)=\frac{(-1)^{n+1}}{n^2}\left(\frac{3I}{\pi^2}\right)^{2/3}$ if $n\ne 0$.

\subsection{Analysis of second order resonances}
\label{classical}

We will focus our analysis on resonances of $(2s+1):2$ type where $s$ is integer. In this case, a perfectly resonant particle moves with a period $T_s = \frac{2s+1}{2}T$, where $T=\frac{2\pi}{\omega}$ is the period of the mirror oscillations, and bounces off the mirror once when it is in the uppermost position and once when it is in the lowermost position and so on. A particle close to the resonance will return to the vicinity of its former position in the phase space after a period $T_s$, but in general there will be a small change in that position. The motion close to the resonance can be described with the help of an effective Hamiltonian, which we will now derive using an original approach.

A perfectly resonant particle hits the mirror alternately in its extreme positions. Let us assume that for $t=0$ the mirror is in its lowest position. Using simple kinematics one can calculate the velocity $v_s$ needed for the particle to be reflected at $t=0$ and then again at $t=T_s$. The mirror's movement has to be taken into account. The resonant value of the action can then be calculated using Eq.~(\ref{Iformula}) and it reads
\be
I_s = \frac{1}{3\pi\omega^3} \left(\frac{(2s+1)\pi}{2}+\frac{2\lambda}{(2s+1)\pi}\right)^3.
\label{Is}
\ee
Let us denote the velocity of a slightly non-resonant particle just before the $j$-th reflection by the mirror by
\be
v_j = v_s + \Delta v_j,
\ee
where $v_s$ is the velocity of a perfectly resonant particle. We also define $\tau_j$ as the amount of time elapsing between $t = jT_s$ and the \textit{j}-th reflection of the non-resonant particle, which in general is not zero. Alternatively, we can say that the quantity $\tau_j$ expresses how much a particle lags behind the resonant trajectory for time $t = jT_s$. 

\begin{figure}[t]
\includegraphics[width=1.\columnwidth]{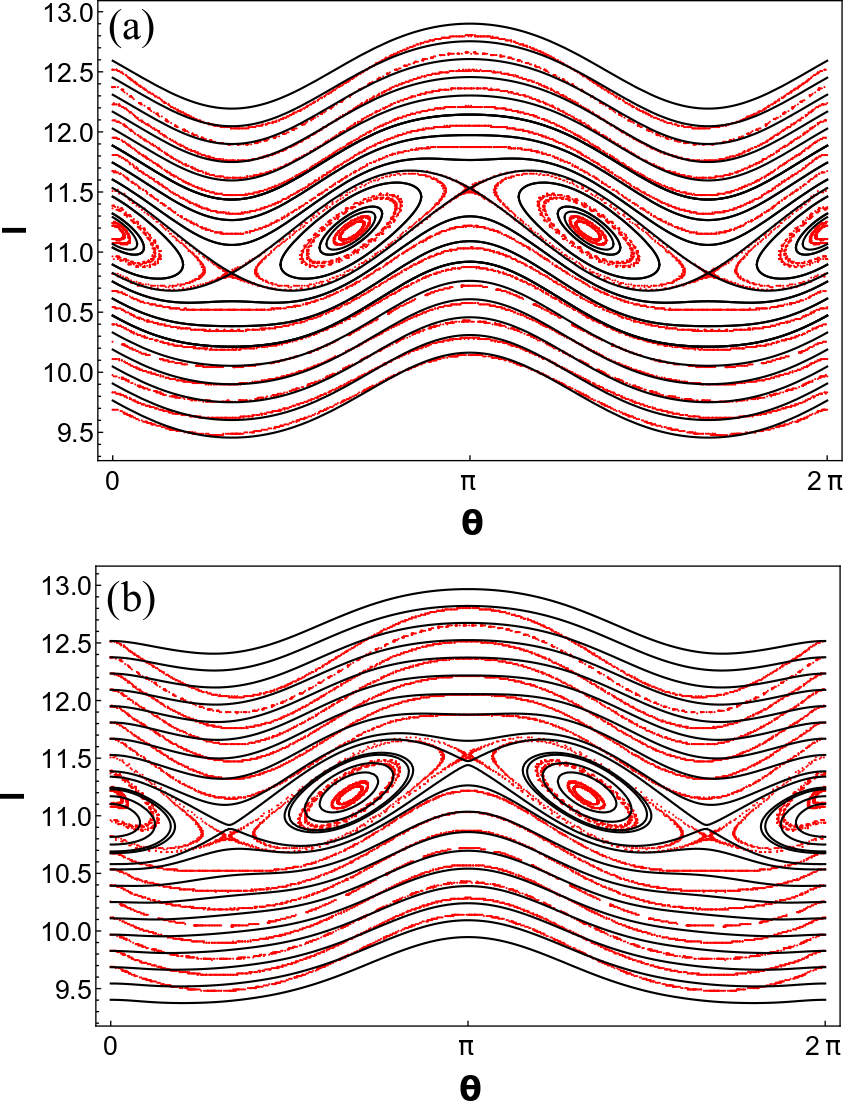}       
\caption{Particle bouncing on the mirror that oscillates with the frequency $\omega=1$ and with the amplitude $\lambda/\omega^2=0.05$. A fragment of the phase space is shown that corresponds to the vicinity of the $3:2$ resonance between the mirror oscillations and the particle motion. Black curves are related to the predictions of the effective Hamiltonian (\ref{heff}) [panel (a)] and to the predictions of the Hamiltonian (\ref{heff2}) [panel (b)]. Red dots in both panels correspond to the stroboscopic picture of the phase space resulting from numerical integration of the exact classical equations of motion generated by the Hamiltonian (\ref{h}). The stroboscopic picture was obtained by collecting $\theta(t)$ and $I(t)$ of classical trajectories every second period of a resonant particle's motion, i.e. every $2T_s=3T$. The gravitational units are used \cite{Giergiel2018a}.}
\label{methods}   
\end{figure}

 Let us calculate how $\Delta v$ and $\tau$ change after consecutive bounces.
A change in the absolute value of $v$ after an elastic reflection is equal to twice the velocity of the mirror. If the first bounce happens when the mirror is in its lowermost position,  
\be
(\Delta v)_1 = (\Delta v)_0 + 2\frac{\lambda}{\omega}\sin\omega\tau_0,
\label{Eqtns1}
\ee
and the time $\Delta t$ between two bounces is
\be
\Delta t = T_s + 2(\Delta v)_1,
\label{period}
\ee
which follows from simple kinematics of a particle in the gravitational field. Thus, the time difference between the next reflection of the strictly resonant particle and the slightly off-resonant one is
\be
\tau_1=\tau_0 + 2(\Delta v)_1.
\label{Eqtns2}
\ee 
In this derivation we have neglected small changes of the mirror's position during the short period $\tau_1$.
Similarly, the next reflection takes place when the mirror is in its highest position and we obtain
\be
(\Delta v)_2 = (\Delta v)_1 - 2\frac{\lambda}{\omega}\sin\omega\tau_1,
\ee
\be
\tau_2=\tau_1 + 2(\Delta v)_2.
\ee
The difference of the action $I$ with respect to the resonant value can be easily expressed in terms of $\Delta v$ using Eq.~(\ref{Iformula}) since at the moment of reflection ($z\approx0$) the potential energy vanishes,
\be
\Delta I \equiv \left.(I-I_s) \approx \left(\frac{\partial I}{\partial p} \right)\right\vert_{I = I_s} \Delta v \approx \left(\frac{2s+1}{2\omega}\right)^2\pi \Delta v.
\ee
In the vicinity of the resonance, the change of $\theta$ by $2\pi$ corresponds to a time period of $T_s$. Thus we can express $\theta$ at time $jT_s$ as
\be
\theta_j = -2\pi\frac{\tau_j}{T_s}.
\label{Theta}
\ee
Using Eq.~(\ref{Eqtns1}) and Eqs.~(\ref{Eqtns2})-(\ref{Theta}) it is possible to calculate changes in $I$ and $\theta$ after two bounces and then approximate the time derivatives
\bea
\dot{\theta}&=&\frac{d\theta}{dt} \approx \frac{\theta_2-\theta_0}{2T_s}, \\
\dot{(\Delta I)}&=&\frac{dI}{dt} \approx \frac{(\Delta I)_2-(\Delta I)_0}{2T_s}.
\eea
When we restrict ourselves to the first non-vanishing order in $\lambda$ and $\Delta I$, we get
\bea
\dot{\theta} &=& -\frac{16\omega^4}{(2s+1)^4\pi^2}\Delta I + \frac{4\lambda\omega}{(2s+1)^2\pi} \sin\left(\frac{2s+1}{2}\theta\right),
\label{thetaprim}
\cr && \\
\dot{(\Delta I)} &=& -\frac{2\lambda\omega}{(2s+1)\pi}\cos\left(\frac{2s+1}{2}\theta\right)\Delta I \cr && + \frac{\lambda^2(2s+1)}{2\omega^2} \sin\left[(2s+1)\theta\right].
\label{iprim}
\eea
Equations~(\ref{thetaprim})-(\ref{iprim}) are actually the Hamilton's equations generated by the following effective Hamiltonian
\bea
H_{eff} &=& -\frac{1}{2}\left[A_s\Delta I-\frac{\lambda}{\omega}\sin\left(\frac{2s+1}{2}\theta\right)\right]^2 \cr&& + \frac{\lambda^2}{4\omega^2}\cos\left[(2s+1)\theta\right], 
\label{heff}
\eea
where
\be
A_s = \frac{4\omega^2}{(2s+1)^2\pi}. 
\label{as}
\ee
Note that the effective mass $m_{\rm eff}$ of a particle described by the effective Hamiltonian is negative, i.e. $m_{\rm eff}=-1/A_s^2$.
The method we have used to derive (\ref{heff}) indicates that the effective Hamiltonian describes the stroboscopic phase space of the problem that can be obtained by plotting the position and momentum of a particle every $2T_s$. In Fig.~\ref{methods}(a) predictions of Eq.~(\ref{heff}) are compared with the results of a numerical integration of the full equations of motion of a particle in the case of the $3:2$ resonance ($s=1$).

There are other methods of calculating an approximate Hamiltonian that describes motion of a particle close to a resonance orbit. The Lie method \cite{Lichtenberg1992} is an elegant approach and it is described in Appendix \ref{appendixlie}. Using the second order Lie method we obtain a different formula, as compared to Eq.~(\ref{heff}), for the desired effective Hamiltonian,
\bea
\widetilde{H}_{eff} &=& -\frac{1}{2}\left[{A}_s\Delta I-\frac{\lambda}{\omega}{B}_s(\theta)\right]^2 
\cr && 
+ \frac{\lambda^2}{\omega^2}{C}_s\cos\left[(2s+1)\theta\right] + {D}_s(\lambda),
\label{heff2}
\eea
where ${A}_s$ is given in (\ref{as}) and
\bea 
{B}_s(\theta) &=& \frac{4(2s+1)}{\pi}\sum_{n=1}^{\infty} \frac{\cos(n\theta)}{n(4n^2-(2s+1)^2)},
\eea
\bea
{C}_s &=& \sum_{n=1}^{2s} \frac{[n-3(2n-2s-1)](2s+1)^2}{n(n-2s-1)^2(2n-2s-1)^2\pi^2} \cr 
&-&  \sum_{n=1}^{\infty}\frac{2[(2n+2s+1)^2+n(n+2s+1)](2s+1)^2}{n^2(2n+2s+1)^2(n+2s+1)^2\pi^2}, 
\cr &&
\label{stalac}
\eea
\bea
{D}_s(\lambda) &=&- \frac{\lambda^2}{\omega^2}\sum_{n=1}^{\infty}\frac{[40n^2-6(2s+1)^2](2s+1)^2}{n^2(2n-2s+1)^2(2n+2s+1)^2\pi^2} \cr &&
 +\frac{\pi^2(2s+1)^2}{4\omega^2}.
  \label{stalad} 
\eea
Comparison of predictions of the Hamiltonian (\ref{heff2}) with numerical results is presented in Fig.~\ref{methods}(b). One can see that the second order Lie method leads not only to a more complicated formula for the effective Hamiltonian but also to less accurate results than the Hamiltonian (\ref{heff}) obtained in a kinematic consideration of a particle bouncing on the oscillating mirror.

In the following we employ quantized versions of the effective Hamiltonian (\ref{heff}) and (\ref{heff2}) in order to find suitable parameters for time crystal behavior and to predict quasi-energy levels corresponding to the resonantly driven particle. The prediction allows us to identify the desired Floquet eigenstates in the numerical diagonalization of the full single-particle Floquet Hamiltonian.

\subsection{Quantum description of second-order resonances}
\label{quantumsingle}

\begin{figure*}[t]
\includegraphics[width=1.8\columnwidth]{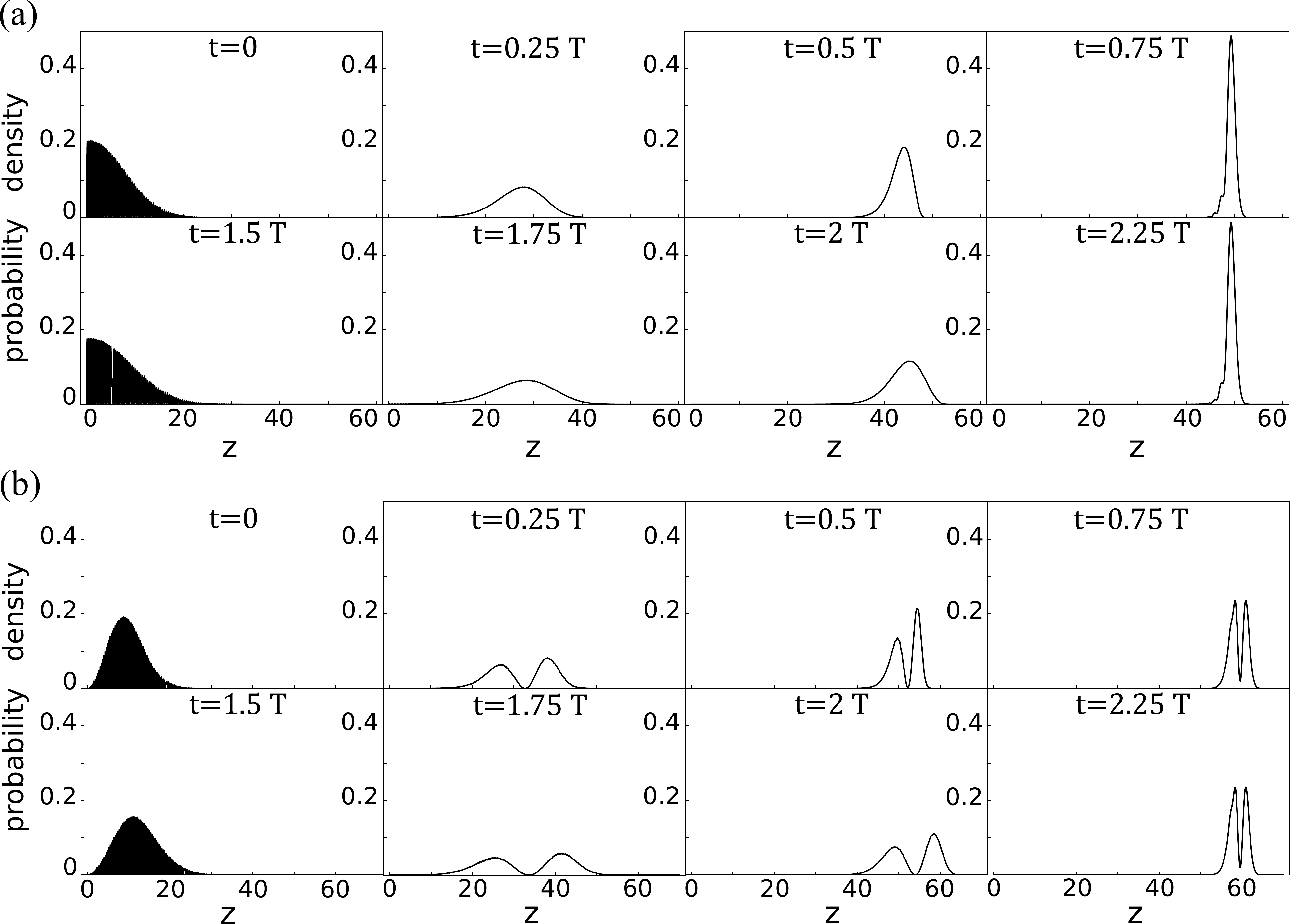}       
\caption{Evolution of Wannier-like wavepackets corresponding to the $3:2$ classical resonance. Wavepackets visit the mirror (which is located at $z=0$) every $3T/2$ like a classical particle on the $3:2$ resonant trajectory, but strictly speaking they evolve with the period $3T$. The reason for this is that the mirror is alternately at different extreme positions at the moments of the reflections. However, when the ratio of the amplitude of the mirror oscillations and the amplitude of the motion of the wavepackets tends to zero ($\lambda\rightarrow 0$), the wavepackets become periodic with the period $3T/2$. Panel (a) illustrates the motion of a wavepacket corresponding to the eigenstates of (\ref{heff}) localized at the bottom of the resonance islands (cf. Fig.~\ref{methods}) for $\omega = 0.47168$ and $\lambda = 0.0825$. Panel (b) shows the motion of a wavepacket related to the first excited eigenstates in the resonance islands for $\omega = 0.4299$ and $\lambda = 0.12$. The apparently black regions are in fact densely packed with interference fringes which result from a superposition of the incoming and reflected parts of the wavepackets. The gravitational units are used \cite{Giergiel2018a}.}
\label{wavepackets}   
\end{figure*} 

The Hamiltonian (\ref{h}) depends explicitly on time, which means that the energy of the system is not conserved. However, since the Hamiltonian is periodic in time, $H(t+T) = H(t)$, we can look for so-called Floquet states which evolve in time with the period $T$. Floquet states are eigenstates of the Floquet Hamiltonian,
\be
\mathcal{H} = H - i\partial_t,
\label{floquet}
\ee
and by the Floquet theorem \cite{Shirley1965} they form a complete basis in the Hilbert space of a particle at any time. Thus, they play a role analogous to Bloch states of a particle in a space-periodic potential. The eigenvalues of $\mathcal{H}$ are called quasienergies and they form a periodic spectrum with the period $\omega=2\pi/T$.

Let us perform canonical quantization of the effective Hamiltonian (\ref{heff}) [or (\ref{heff2})], i.e. $\theta \rightarrow \hat{\theta}$ and $\Delta I \rightarrow -i\frac{\partial}{\partial \theta}$, and calculate its eigenstates \cite{Giergiel2018a}. The effective mass of a particle in (\ref{heff}) is negative, thus, the highest energy eigenstates of (\ref{heff}) are localized (in a semiclassical sense) in the resonant elliptical islands visible in Fig.~\ref{methods} if the islands are sufficiently big \cite{Buchleitner2002}. The size of the islands depends on $\lambda$ and $\omega$. The parameter $\lambda$ cannot be too big because then the effective Hamiltonian is not valid. However, for small $\lambda$ and sufficiently small $\omega$ (or equivalently large $I_s$) the islands are big enough to host a few quantum eigenstates. Semiclassical approach (see Appendix~\ref{appndxsemiclas}) allows one to estimate the number of eigenstates trapped inside the islands 
\be
n _{\rm trapped} \approx \lambda \frac{\pi (2s+1)}{4\sqrt{2}\omega^3} \approx \lambda \frac{1.4 I_s}{ (2s+1)^2}.
\label{ntrapped}
\ee
The eigenstates are represented by superpositions of localized wavepackets that evolve along the resonant orbit when we plot them in the laboratory frame and they are actually the Floquet states of the system.

In the case of the $3:2$ resonance, there are three elliptical islands (see Fig.~\ref{methods}) and therefore there are three Floquet states $\phi_i(z,t)$ corresponding to three eigenstates localized at the bottom of the elliptical islands. 
The corresponding three energy levels of (\ref{heff}) are nearly degenerate,
but there is a tiny splitting in the energies which is related to a tunneling process. That is, a superposition of the three eigenstates allows one to extract a single localized wavepacket that, in the laboratory frame, evolves along the resonant orbit and slowly tunnels to the neighboring islands. It is possible to extract three such localized wavepackets $w_j(z,t)$ by superposing the Floquet states $\phi_i(z,t)$,
\be
\begin{pmatrix}
w_1 \\ w_2 \\ w_3
\end{pmatrix}
 = \frac{1}{\sqrt{3}}
 \begin{pmatrix}
  1 & 1 & 1 \\
  1 & e^{i\frac{2\pi}{3}} & e^{-i\frac{2\pi}{3}} \\
  1  & e^{-i\frac{2\pi}{3}}  & e^{i\frac{2\pi}{3}}
 \end{pmatrix}
 \begin{pmatrix}
 \phi_1 \\ e^{-i\frac{\omega}{3}t}\phi_2 \\ e^{i\frac{\omega}{3}t}\phi_3
 \end{pmatrix}.
\ee
The functions $w_j(z,t)$ are periodic with the period $3T$ --- in Fig.~\ref{wavepackets}(a) we present one of them at different moments of time. However, when we take one of the functions $w_j$ as an initial state of a particle and evolve it according to the Schr\"odinger equation we observe that on a long time scale a particle tunnels to the other two wavepackets which are initially unoccupied. It is apparent when we restrict to the Hilbert subspace spanned by the three periodic functions $w_j(z,t)$, i.e. the wavefunction of a particle is assumed to be $\psi(z,t)=\sum_{j=1}^3a_j(t)w_j(z,t)$, then the Floquet Hamiltonian in this subspace takes a form of the tight-binding model \cite{sacha15a,Giergiel2018a}. That is, the quasi-energy of a particle in the subspace reads 
\bea
E
&\approx&-\frac12\sum\limits_{i\ne j}J_{ij}\;a_i^*a_j,
\label{tbm}
\eea
with 
\be
J_{ij} = -\frac{2}{3T}\int\limits_0^{3T} dt \int\limits_0^\infty dz\;w_i^*(z,t)\;\mathcal{H}\;w_j(z,t).
\label{j}
\ee
In (\ref{j}) $J_{ij}$'s (with the same modulus $J=|J_{ij}|$) are amplitudes which describe a tunneling of a particle. For example, when the initial wavefunction of a particle is $\psi(z,0)=w_2(z,0)$, the time evolution according to the tight-binding model (\ref{tbm}) leads to $\psi(z,t)=\sum_{j=1}^3a_j(t)w_j(z,t)$ and $a_2(t)=0$ at $t\propto1/J$ which means that the particle has tunneled out to neighboring wavepackets $w_1$ and $w_3$. 

This approach can be generalized to any $(2s+1):2$ resonance. When $s\rightarrow\infty$, the eigenvalues of the tight-binding model form an energy band of the width of $J$ and the time-periodic localized wavepackets $w_j(z,t)$ play a role of Wannier states known in condensed matter physics \cite{Dutta2015}. The wavepackets are strictly periodic with the period $(2s+1)T$ but they revisit the mirror every period $(2s+1)T/2$ because they evolve along the classical $(2s+1):2$ resonant orbit. Due to the fact that they bounce off the mirror alternately when the mirror is at the uppermost and lowermost positions, the wavepackets are not perfectly periodic with the period $(2s+1)T/2$, see Fig.~\ref{wavepackets}. However, the imperfection disappears when the amplitude of the mirror oscillations is small as compared to the amplitude of a particle motion, i.e. when $\lambda/(2s+1)^2\rightarrow0$. For a given $s$, when $\lambda\rightarrow0$ we have to ensure that the elliptical islands visible in Fig.~\ref{methods} are sufficiently large to host quantum states. It is not a problem because when $\lambda$ goes to zero, we can decrease the frequency $\omega$ of the mirror oscillations so that $\lambda/\omega^3=$constant and the number $n _{\rm trapped}$ of quantum states trapped in the islands remains intact, see (\ref{ntrapped}). In such a limit the wavepackets $w_j$ become periodic with the period $(2s+1)T/2$.

We have analyzed eigenstates of (\ref{heff}) localized at the bottom of the $3:2$ resonant islands. Similarly one can define a Hilbert subspace spanned by the three excited eigenstates in the islands. The corresponding Floquet states are superpositions of three localized wavepackets, which we denote by $\tilde w_j(z,t)$, but the wavepackets possess different shape as compared to the eigenstates localized at the bottom of the resonant islands. Indeed, the excitation creates nodes in the wavefunctions and consequently the density profiles of the wavepackets reveal a hole --- in Fig.~\ref{wavepackets}(b) we present one of the them at different moments of time. The wavepackets can be chosen as basis vectors that span a Hilbert subspace and within this subspace quasi-energy of a particle is again given by the tight-binding model like in Eq.~(\ref{tbm}) with analogous tunneling amplitudes, i.e. $J_{ij}\rightarrow \tilde J_{ij}$ where $w_j(z,t)\rightarrow\tilde w_j(z,t)$ in (\ref{j}). When we consider a general $(2s+1):2$ resonance and when $s\rightarrow\infty$, the eigenvalues of the effective Hamiltonian (\ref{heff}) corresponding to this subspace form an energy band of the width $\tilde J=|\tilde J_{ij}|$. This band, which in the context of ultra-cold atoms in an optical lattice is called $p$-band \cite{Dutta2015}, and the previous band are separated in energy by the gap which can be much larger than their widths $J$ and $\tilde J$.

In the following we will consider a many-body system of interacting ultra-cold atoms which are bouncing resonantly on an oscillating atom mirror. If the interaction energy per particle is much smaller than the energy gap between the bands that we have just defined, a description of the resonantly driven many-body system can be restricted to one of the bands and one obtains effectively the Bose-Hubbard model \cite{sacha15a,Giergiel2018a}, i.e. a many-body generalization of the tight-binding model (\ref{tbm}).

\section{Many-body system and fractional time crystal formation}
\label{sec2}

We consider an $N$-body system that consists of interacting ultra-cold atoms bouncing on an oscillating atom mirror. If the bounces are resonant with the mirror motion, one can reduce description of the system to one of the energy bands which we have identified in the previous section. In the present section we show that the lowest energy states within a band can reveal spontaneous breaking of the time translation symmetry of the Hamiltonian and start moving with a period different from the mirror oscillation period $T$ \cite{Sacha2015}. In the limit of small amplitude of the mirror oscillations, the symmetry broken states evolve with the period $(2s+1)T/2$ demonstrating that a new class of discrete time crystals can be realized in an experiment, i.e. time crystals evolving with rational multiples of a driving period. While our approach is valid for any $(2s+1):2$ resonance, here we will focus on the $3:2$ case.

The entire Hilbert space of the $N$-body system is very large. However, if we are interested in the $3:2$ resonant bouncing of an atomic cloud on an oscillating mirror we may restrict ourselves to a subspace spanned by Fock states $|n_1,n_2,n_3\ra$ where $n_j$'s are numbers of atoms that occupy the Wannier-like wavepackets $w_j(z,t)$. The latter are obtained by superposing the eigenstates localized at the bottom of the elliptical islands, see Sec.~\ref{quantumsingle}. Then, the Floquet many-body Hamiltonian can be approximated as follows \cite{sacha15a,Giergiel2018a}
\bea
\hat{\cal H}&=&\frac{1}{3T}\int\limits_0^{3T} dt \int\limits_0^\infty dz \;\hat{\psi}^\dagger\left(H + \frac{g_0}{2}\hat{\psi}^\dagger\hat{\psi}-i\partial_t\right)\hat{\psi} \cr
&\approx& -\frac12\sum_{i\neq j} J_{ij}\hat a_i^\dagger \hat a_j + \frac{g_0}{2}\sum_{i,j=1}^{3} U_{ij} \hat a_i^\dagger \hat a_j^\dagger \hat a_j \hat a_i.
\label{bhh}
\eea
In (\ref{bhh}) we have truncated the bosonic field operator $\hat\psi(z,t)\approx\sum_{i=1}^3w_i(z,t)\hat a_i$ where $\hat a_i$'s are the standard bosonic annihilation operators, and 
\bea
U_{ij} = \frac{2}{3T}\int\limits_0^{3T} dt \int\limits_0^\infty dz~|w_i|^2|w_j|^2, 
\eea
for $i\ne j$ and similar $U_{ii}$ but by the factor 2 smaller.
The many-body effective Hamiltonian (\ref{bhh}) is the Bose-Hubbard model and it is valid provided the interaction energy per particle is much smaller than the energy gap between the energies of the eigenstates of (\ref{heff}) localized at the bottom of the elliptical islands and the energies of excited eigenstates inside the islands, see Sec.~\ref{quantumsingle}. In the following we consider examples where the interaction energy per particle, $g_0N U_{ii}$, is of the order of $J$ while the energy gap is of the order of $10^3J$.

\begin{figure}[t]
\includegraphics[width=1.1\columnwidth]{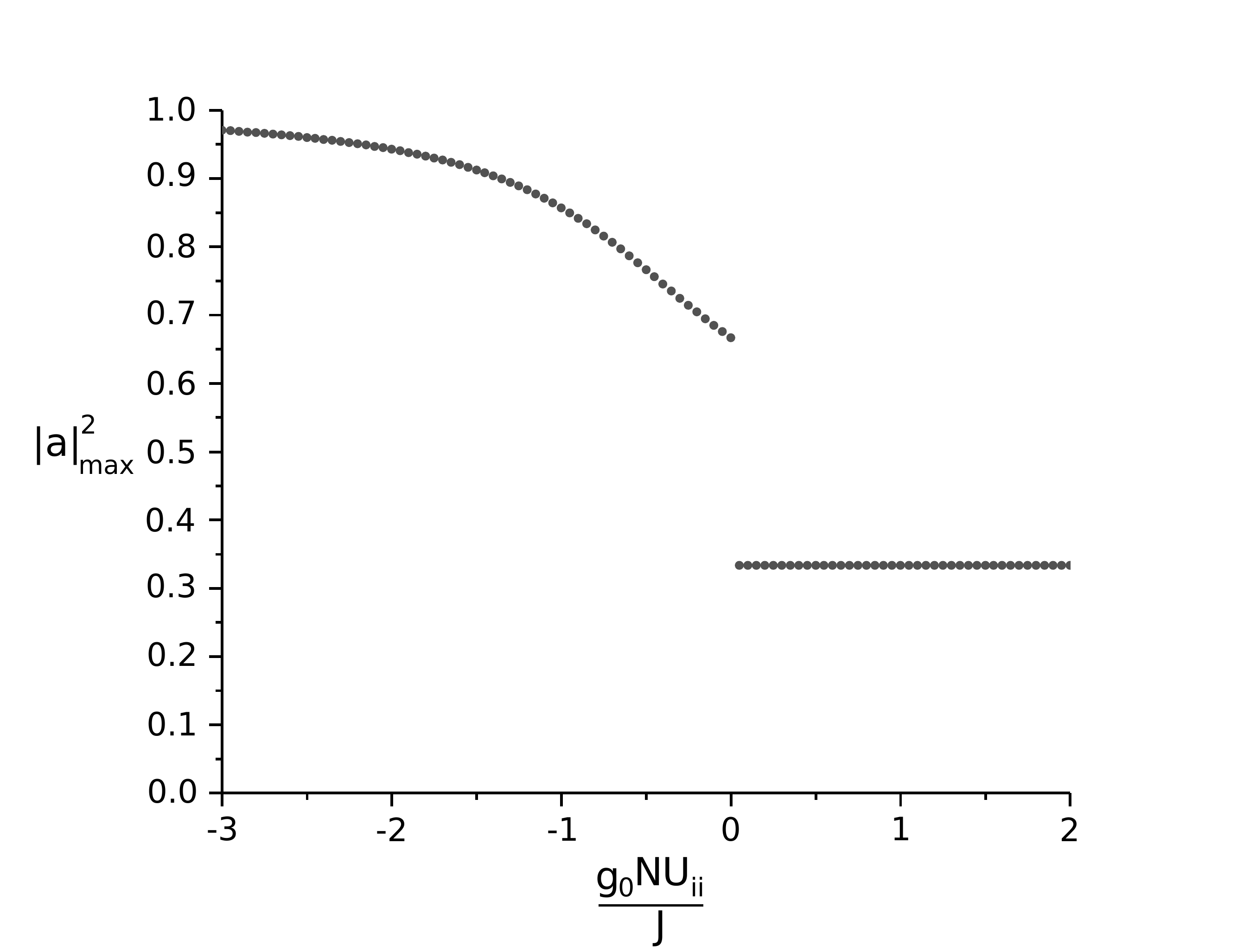}       
\caption{Spontaneous breaking of time translation symmetry in the system of ultra-cold atoms bouncing resonantly on an oscillating atom mirror --- the $3:2$ resonant condition is fulfilled. $|a|_{max}$ is the absolute value of the dominant coefficient $a_j$ in the expansion of the state, $\psi=\sum_{j=1}^3a_jw_j$, which minimizes the energy (\ref{emean}). $|a|^2_{max}=\frac{1}{3}$ corresponds to the symmetry preserving state, while $|a|_{max}^2=1$ to a single Wannier-like wavepacket which evolves with a period different from the driving period. In the manuscript we focus on the parameters where all tunneling amplitudes are real, $J_{ij}=|J_{ij}|=J$. Then, any attractive interactions between particles break the time translation symmetry because for $g_0=0$, the ground state energy level of (\ref{emean}) is degenerate.}
\label{graph}   
\end{figure} 

For weak repulsive interactions, the ground state of (\ref{bhh}), i.e. the lowest energy state of the system in the subspace we consider, is a Bose-Einstein condensate $\Psi_0(z_1,\dots,z_N,t)=\prod_{i=1}^N\phi_1(z_i,t)$ where all atoms occupy the same single-particle wavefunction which is a Floquet state $\phi_1(z,t)$ that evolves with the period $T$ of the mirror oscillations. However, when the interactions are attractive ($g_0<0$) and sufficiently strong it becomes energetically favorable to group all atoms in a single localized wavepacket $w_j$ and form a Bose-Einstein condensate $\prod_{i=1}^Nw_j(z_i,t)$ because it decreases the energy of the system. However, such a state evolves with a period different from $T$ and it cannot be a Floquet many-body state because it breaks the time translation symmetry of the Hamiltonian. In order to reconcile the energy and symmetry requirements, the ground state is a superposition of Bose-Einstein condensates, $\Psi_0\propto\sum_{j=1}^3\prod_{i=1}^Nw_j(z_i,t)$. Such a state evolves with the period $T$ despite the fact that $w_j(z,t)$'s are periodic with the period $3T$ because after every period $T$, $w_j(z,t)$'s exchange their role. Note that the ground state $\Psi_0$ is a macroscopic superposition and it is sufficient to perform a destructive measurement of the position of a single atom and the state collapses to one of the Bose-Einstein condensates \cite{Sacha2015}, $\Psi_0\rightarrow \Psi=\prod_{i=1}^{N-1}w_j(z_i,t)$ --- which condensate is realized depends on the result of the measurement. The collapse is an example of a spontaneous symmetry breaking process responsible for time crystal formation which can happen due to an intentional measurement or an atom loss or some other perturbation. In the limit when $N\rightarrow\infty$ but $g_0N=$constant, the symmetry-broken state $\Psi(z_1,\dots,z_{N-1},t)$ evolves with a period different from the driving period $T$ and it does not decay because the macroscopic tunneling of a condensate takes an infinite time \cite{Sacha2015}.

We would like to stress that the periodically evolving Bose-Einstein condensate that demonstrates breaking of the time translation symmetry cannot be achieved by cooling a thermal cloud of atoms in the presence of the time-periodic driving. The ground state of the Bose-Hubbard model (\ref{bhh}) does not correspond to the ground state of the original periodically driven system. Actually there is no ground state of the periodically driven system because the quasi-energy spectrum is unbounded and periodic. In order to demonstrate the time crystal behavior, first a Bose-Einstein condensate has to be prepared in a trap and then released from the trap and loaded into the classical resonant orbit. Such a scenario is analyzed in details in Ref.~\cite{Giergiel2018a} --- see also the last paragraphs of the present Section.

Our analysis indicates that the many-body system, both when the interactions are repulsive  and when they are attractive and a time crystal forms, is actually a Bose-Einstein condensate and the mean-field approach may be applied \cite{Pethick2002}. The easiest way to switch to the mean-field description relies on the substitution $\hat a_i\rightarrow\sqrt{N}a_i$ where $a_i$'s are complex numbers. Then, the energy of the system per particle reads
\be
E=-\frac12\sum_{i\neq j} J_{ij}a_i^* a_j + \frac{g_0N}{2}\sum_{i,j=1}^{3} U_{ij} |a_i|^2|a_j|^2.
\label{emean}
\ee
For an appropriate choice of $\omega$ and $\lambda$, the tunneling amplitudes can be made real $J_{ij}=|J_{ij}|=J$. For example, for $\lambda = 0.0825$ and $\omega = 0.47168$ we obtain $J\approx 4.9\times10^{-5}$. Because of the symmetry of the Wannier-like localized wavepackets $w_j$, the on-site interaction coefficients do not depend on the index $i$ and for the chosen parameters $U_{ii}=0.12$. Similarly it is the case for the long-range interaction coefficients which are an order of magnitude smaller than the on-site ones, i.e. $U_{ij}=0.024$ for $i\ne j$. If the interactions are repulsive ($g_0>0$), the lowest value of the energy (\ref{emean}) corresponds to the state with uniform superposition of the localized wavepackets $w_j$, i.e. $a_1=a_2=a_3=1/\sqrt{3}$. Such a mean-field solution describes a Bose-Einstein condensate and evolves with the period of the mirror oscillations $T$. However, when the interactions are attractive $g_0<0$, the mean-field approach reveals spontaneous breaking of the discrete time translation symmetry of the many-body Hamiltonian because the lowest energy states are degenerate and each of them evolves with a period different from $T$. If $g_0NU_{ii}/J\lesssim -2$ the lowest energy solutions reduce practically to single localized wavepackets $w_j$, see Fig.~\ref{graph}. 

In the case when the tunneling amplitudes $J_{ij}$ are real and positive, an analytical expression for the lowest energy level can be found for $g_0<0$. This expression is lengthy and not crucial for our considerations, so we only present its asymptotic forms. Let us denote the quantity $\frac{2g_0N(U_{ii}-U_{ij})}{J}$ by $\kappa$. For $g_0N \rightarrow 0_-$, one of the coefficients $a_j$ of the ground state solution, $\psi=\sum_{j=1}^3a_jw_j$, takes the value $\sqrt{\frac{2}{3}} - \frac{1}{18}\sqrt{\frac{2}{3}}\kappa$ and the other two take the value $-\sqrt{\frac{1}{6}} - \frac{1}{18}\sqrt{\frac{2}{3}}\kappa$. For $g_0N \rightarrow -\infty$, the dominant coefficient is equal to $1 - \frac{1}{\kappa^2}$ and the other two $\frac{1}{\kappa}$. 

Strictly speaking all the symmetry-broken solutions evolve with the period $3T$. However, ultra-cold atoms actually revisit the mirror with the period $3T/2$ and it is only due to the fact that the mirror is alternately in the uppermost and lowermost positions at the moments of the bouncings that the evolution is not perfectly periodic with the period $3T/2$. This imperfection disappears when the ratio of the amplitude of the mirror oscillations and the amplitude of the motion of atoms tends to zero, i.e. $\lambda\rightarrow0$. In Fig.~\ref{wavepackets}(a) we show an example where one can see that even if the resonant value of the action $I_s$ is not very big and consequently the amplitude of the atom motion not very large, it is already hard to identify differences between consecutive bounces. If we chose smaller $\omega$ it would not be possible to distinguish differences between consecutive bounces. Thus, in our system the time translational symmetry of the Hamiltonian is spontaneously broken and a time crystal emerges which evolves with the period $3T/2$ which is a rational multiple of the driving period.

The same analysis can be performed when ultra-cold atoms are prepared in the Hilbert subspace related to the excited eigenstates in the resonance islands, i.e. when we look for the lowest energy solution of the Bose-Hubbard model (\ref{emean}) with $J_{ij}\rightarrow \tilde J_{ij}$ in the form $\psi(z,t)=\sum_{j=1}^3a_j\tilde w_j(z,t)$, see Sec.~\ref{quantumsingle}. For the parameters used in Fig.~\ref{wavepackets}(b) and for $g_0NU_{ii}/J\lesssim -2$ we obtain that the lowest energy state become nearly single localized wavepackets $\tilde w_j$, i.e. $|\la w_j|\psi\ra|^2>0.95$. Thus, time translation symmetry is spontaneously broken and the system evolves with the period $3T/2$ if the amplitude of the mirror oscillations is small as compared to the amplitude of the bounces. The difference with respect to the previous case is that now the density profile of the wavepackets $\tilde w_j$ reveals a hole at the center, see Fig.~\ref{wavepackets}.

The experiment demonstrating the discrete time crystals evolving with rational multiples of the driving period that we consider here can be realized if a Bose-Einstein condensate is prepared in a trap above the mirror at the classical turning point. Then, the release of the atom cloud from the trap at a proper moment of time synchronized with the mirror oscillations allows one to load atoms to a classical resonant orbit \cite{Giergiel2018a}. If the shape of the cloud is adjusted to the shape of a localized wavepcket $w_j$ at the position of the classical turning point, it means that our system is prepared in a single Wannier-like wavefunction of the Bose-Hubbard model (\ref{bhh}) and it will evolve with the period $(2s+1)T/2$ if we are in the time crystal regime otherwise it will decay, i.e. atoms will tunnel to neighboring wavepackets on a time scale proportional to $1/J$. 

In order to demonstrate time crystal behavior where atoms occupy an excited wavepacket $\tilde w_j$, one has to prepare initially a Bose-Einstein condensate in an excited state of a trap, i.e. in a state which for repulsive interaction corresponds to a dark soliton \cite{Pethick2002}. Next, we can apply the same procedures of loading an atom cloud to a resonant orbit and detection of time crystal behavior as in the previous case.

\section{Summary and conclusions}
\label{seccon}

We have considered ultra-cold atoms bouncing on an oscillating atom mirror and analyzed a class of second-order resonances between the atom motion and the mirror oscillations. By means of an original approach we have derived an effective single-particle Hamiltonian which allows us to analyze the Floquet states that describe the quantum resonant evolution of a single atom. Then, we have switched to the many-body case and derived an effective Bose-Hubbard model which describes quantum many-body resonant dynamics.

We have shown that for sufficiently strong attractive interactions between atoms a spontaneous symmetry breaking of the time translation symmetry of the Hamiltonian occurs and the lowest energy states within a resonant Hilbert subspace of the system approach single localized wavepackets, which in the limit $\lambda \rightarrow 0$ evolve periodically with a fractional multiple of the driving period. Such a system constitutes a different class of time crystals as compared to those already demonstared in the laboratory and it can be realized experimentally by means of a cloud of ultracold atoms bouncing on an oscillating atomic mirror \cite{Steane95, Giergiel2018a}. 

We have focused on fractional time crystals where the ratio of periods of their motion and the driving period equals $\frac{2s+1}{2}$ with integer $s$. However, other rational numbers can be realized if one considers higher order resonances in the system.



\section*{Acknowledgments}
We thank Peter Hannaford for reading of the manuscript and suggestion of its title. Support of the National Science Centre, Poland via Project No. 2016/21/B/ST2/01095 (P.M.) and QuantERA programme No.~2017/25/Z/ST2/03027 (K.S.) is acknowledged. 

\appendix
\section{Lie method}
\label{appendixlie}
Our description of the Lie method is based on \cite{Lichtenberg1992}. Here we will only describe its application up to the second order in a perturbation parameter $\lambda$.

The goal is to transform an original Hamiltonian $H(x_i,p_i)$ to a new one $\bar{H}(\bar{x}_i,\bar{p}_i)$, which approximates the motion in a resonance island. The old variables are transformed into the new ones by an operator $\mathcal{T}$:
\begin{equation}
\bar{x}_i = \mathcal{T}x_i, \quad \bar{p}_i = \mathcal{T}p_i. 
\label{transformacje}
\end{equation}
$H$, $\bar{H}$ and $\mathcal{T}$ can be expanded in power series in the perturbation parameter $\lambda$:
\begin{equation}
H = \sum_{n=0}^\infty \lambda^n H_n,\quad\bar{H} = \sum_{n=0}^\infty \lambda^n \bar{H}_n,\quad\mathcal{T} = \sum_{n=0}^\infty \lambda^n \mathcal{T}_n.
\end{equation}
It turns out that $\mathcal{T}_n$ can be expressed by operators $L_n$ of the form 
\begin{equation}
L_n= \{w_n(x_i,p_i),\;\cdot\;\},
\end{equation}
where $\{\;\cdot\;,\;\cdot\;\}$ is the Poisson bracket and $w_n$ are some functions that need to be calculated. For $n=0,1,2$:
\bea
\mathcal{T}_0 &=& 1, \cr
\mathcal{T}_1& =& -L_1, \cr
\mathcal{T}_2 &=& -\frac{1}{2}L_2 + \frac{1}{2}L_1^2.
\label{rownnaT}
\eea
Equations for $w_n$ take the forms
\bea
\frac{\partial w_1}{\partial t} + \{w_1, H_0\} &=& \bar{H}_1-H_1, \label{rownnaw1} \\
\frac{\partial w_2}{\partial t} + \{w_2, H_0\}& =& 2(\bar{H}_2-H_2) - \{w_1,\bar{H}_1+H_1\}. \cr&& \label{rownnaw2}
\eea
$\bar{H}_1$, $\bar{H}_2$ etc. are such that the right-hand sides of the Eqs.~(\ref{rownnaw1})-(\ref{rownnaw2}) averaged over a quickly varying variable are 0, i.e.
\bea
\bar{H}_1 &= &\langle H_1 \rangle, \label{usrednianie1} \\
\bar{H}_2 &=& \langle H_2 + \frac{1}{2}\{w_1,\bar{H}_1+H_1\} \rangle. \label{usrednianie2}
\eea
The final result, i.e. the effective Hamiltonian up to the second order in $\lambda$, reads
\begin{equation}
\bar{H} = H_0(\bar{x}_i,\bar{p}_i, t) + \lambda \bar{H}_1(\bar{x}_i,\bar{p}_i, t) + \lambda^2\bar{H}_2(\bar{x}_i,\bar{p}_i, t), \label{wynikLiego}
\end{equation}
where $\bar{x}_i,\bar{p}_i$ can be obtained from (\ref{transformacje}) having calculated $\mathcal{T} = \mathcal{T}_0 + \lambda \mathcal{T}_1 + \lambda^2 \mathcal{T}_2$ from Eq.~(\ref{rownnaT}).

Before we apply the described procedure to our system, we first perform a canonical transformation 
\bea
I' &=& I, \\
\theta' &=& \theta - \frac{2}{2s+1}\omega t,
\label{przejscie}
\\
H'(I', \theta',t) &=& H\left(I', \theta'+ \frac{2}{2s+1}\omega t,t\right) - \frac{2}{2s+1}\omega I'. \cr && \label{noweh}
\eea
This transformation corresponds to passing to the frame of reference rotating with the frequency $\frac{2}{2s+1}\omega$. Here $H$ is given by Eq.~(\ref{h_action}). It means that $H'_1$, that is the part of the Hamiltonian proportional to $\lambda$, is given by the series present in Eq.~(\ref{h_action}) while $H'_2 = 0$. It can be easily checked that with the new Hamiltonian (\ref{noweh}), $\dot{\theta'}$ and $\dot{I'}$ in the vicinity of the resonant action, i.e. for $I'\approx I'_s$ where
\be
I'_s = \frac{1}{3\pi\omega^3} \left(\frac{(2s+1)\pi}{2}\right)^3,
\ee
are of the order of $\lambda$. It means that the only quickly varying variable is the time $t$. It is easy to show that $\bar{H'}_1 = \langle H'_1 \rangle_t = 0$, so the perturbation calculations have to be carried out at least up to the second order in $\lambda$. We assume that $w_1$ is given by the series 
\be
w_1 = \sin \omega t \sum_{n} c_n(I')e^{in\left(\theta'+\frac{2}{2s+1}\omega t\right)}.
\ee
The coefficients $c_n(I')$ are calculated with the help of Eq.~(\ref{rownnaw1}). Using the expression for $w_1$, we obtain the final effective Hamiltonian $\bar{H'} = H'_0 + \bar{H}_2'$ which reads
\be
\bar{H'} = D_s\frac{\lambda^2}{\omega^2} - \frac{8\omega^4}{\pi^2(2s+1)^4} {\Delta \bar I'}^2 + C_s\frac{\lambda^2}{\omega^2}\cos\left[(2s+1)\bar{\theta'}\right], \label{wzorzbarami}
\ee
where
\bea
C_s &=& -\sum_{n=1}^{\infty}\frac{2[(2n+2s+1)^2+n(n+2s+1)](2s+1)^2}{n^2(2n+2s+1)^2(n+2s+1)^2\pi^2} \cr
&&+ \sum_{n=1}^{2s} \frac{[n-3(2n-2s-1)](2s+1)^2}{n(n-2s-1)^2(2n-2s-1)^2\pi^2}, \\
D_s &=& 
- \frac{\lambda^2}{\omega^2}\sum_{n=1}^{\infty}\frac{[40n^2-6(2s+1)^2](2s+1)^2}{n^2(2n-2s+1)^2(2n+2s+1)^2\pi^2}  \cr
&&\frac{\pi^2(2s+1)^2}{4\omega^2}.
\eea
The Hamiltonian (\ref{wzorzbarami}) is written in the transformed coordinates (\ref{transformacje}), thus to compare its predictions with the results of the numerical simulations presented in Fig.~\ref{methods} we have to return to the laboratory frame. As we are interested in the second-order expansion, we can approximate 
\be
\bar{\theta'}\approx\theta'=\theta-\frac{2}{2s+1}\omega t,
\ee
and
\be
{\Delta \bar I'} \approx \Delta I' - \{w_1, \Delta I'\} = \Delta I - \frac{(2s+1)^3\lambda}{4\omega^3}F(\theta, t),
\ee
where
\be
F(\theta, t) = \sum_{n = 1}^{\infty} \left(\frac{\cos \left(n\theta+\omega t\right)}{n^2 \left(2n+2s+1\right)} + \frac{\cos \left(n\theta -\omega t\right)}{n^2 \left(2n-2s-1\right)}\right).
\label{funkcjaf}
\ee
We would like to compare trajectories generated by the Hamiltonian (\ref{wzorzbarami}) with the exact stroboscopic picture of the phase space. Plugging $t = j\frac{2\pi}{\omega}t$, where $j$ is integer, into Eq.~(\ref{funkcjaf}) and combining all previous equations we obtain the effective Hamiltonian in the form of Eq.~(\ref{heff2}).

\section{Semiclassical analysis}
\label{appndxsemiclas}

Before we start discussing semiclassical analysis, let us notice that in Eq.~(\ref{heff}) we can use a simple canonical transformation, ${\Delta \bar I} = \Delta I - \frac{\lambda}{A_s \omega} \sin \left( \frac{2s+1}{2}\theta \right)$ and $\bar{\theta}=\theta$, to obtain the effective Hamiltonian in the form
\be
{H}_{eff} = -\frac{8\omega^4}{\pi^2(2s+1)^4}{\Delta\bar I}^2 + \frac{\lambda^2}{4\omega^2}\cos\left[(2s+1)\bar{\theta}\right] + K_s(\lambda).
\label{heffnew}
\ee 
In (\ref{heffnew}) we have added a constant term $K_s(\lambda)$ which  cannot be obtained in the analysis presented in Sec.~\ref{classical} but can be determined by comparison of the effective Hamiltonians (\ref{heff}) and (\ref{heff2}). This term is irrelevant for classical dynamics but it is important when we want to compare quasi-energy levels obtained by means of the effective Hamiltonians and in the exact diagonalization of the Floquet Hamiltonian, Eqs.~(\ref{h}) and (\ref{floquet}).

In order to calculate quasi-energy levels semiclassically one has to take into account that our system is periodically driven. The quantum Floquet Hamiltonian corresponds to the classical Hamiltonian of the system when the phase space is extended by the time variable $t$ and the conjugate momentum $P_t$ which is the energy taken with the minus sign \cite{Buchleitner2002},
\be
\mathcal{H}_{eff}(\bar{\theta},{\Delta\bar I}, t, P_t) = H_{eff}(\bar{\theta},{\Delta\bar I}) + P_t.
\label{hextended}
\ee

The energy levels can be found by using the WKB method simultaneously for the two pairs of variables, $(\bar{\theta},{\Delta\bar I})$ and $(t, P_t)$ \cite{Buchleitner2002}. From the discussion in Sec.~\ref{classical} it is evident that a particle returns to the vicinity of the initial position in the phase space every $2T_s$. Because $H_{eff}$ does not depend on $t$, the momentum $P_t$ is constant. Hence the quantization condition is
\be
\int_0^{2 T_s} P_t d t = (2s+1)\frac{2\pi}{\omega}P_t = 2\pi j,
\ee
where $j$ is an integer number. Applying this results in Eq.~(\ref{hextended}) we see that the semiclassical quasi-energy spectrum is periodic with the period $\frac{\omega}{2s+1}$. The other quantization condition reads
\be
\oint {\Delta\bar I} d \obar{\theta} = 2 \pi \left(k+\frac{1}{2}\right),
\label{wkbquant}
\ee
where $k$ is integer and the integration is along a closed trajectory in the phase space \cite{Buchleitner2002}. This equality allows one to find quasi-energies of the Floquet states located at the bottom of the classical islands and the excited states in the islands.
Alternatively, one can perform canonical quantization of the effective Hamiltonian,
$\bar{\theta} \rightarrow \hat{\theta}$ and ${\Delta\bar I} \rightarrow \hat{\Delta I} = -i\frac{\partial}{\partial \theta}$, and diagonalize $H_{eff}$. 

Predictions of the quantized versions of the effective Hamiltonians (\ref{heff}) and (\ref{heff2}) can be compared with the results of the exact diagonalization of the full Floquet Hamiltonian of the system.  
For example, for $s=1$, $\lambda=0.12$ and $\omega = 0.43$, the Hamiltonian (\ref{heff2}) provides a value of the energy gap between the eigenstates located at the bottom of the elliptical islands and the first excited eigenstates with a relative accuracy of $10\%$, while for the Hamiltonian (\ref{heff}) the relative accuracy is $2\%$.

Equation~(\ref{wkbquant}) can be used to estimate the number of states trapped in a resonance island. This number is approximately equal to the area of the resonance island in the phase space divided by $2\pi$
\be
n _{trapped} \approx \lambda \frac{\pi (2s+1)}{4\sqrt{2}\omega^3} \approx \lambda \frac{1.4 I_s}{ (2s+1)^2}.
\ee 


\begin{thebibliography}{76}
\expandafter\ifx\csname natexlab\endcsname\relax\def\natexlab#1{#1}\fi
\expandafter\ifx\csname bibnamefont\endcsname\relax
  \def\bibnamefont#1{#1}\fi
\expandafter\ifx\csname bibfnamefont\endcsname\relax
  \def\bibfnamefont#1{#1}\fi
\expandafter\ifx\csname citenamefont\endcsname\relax
  \def\citenamefont#1{#1}\fi
\expandafter\ifx\csname url\endcsname\relax
  \def\url#1{\texttt{#1}}\fi
\expandafter\ifx\csname urlprefix\endcsname\relax\def\urlprefix{URL }\fi
\providecommand{\bibinfo}[2]{#2}
\providecommand{\eprint}[2][]{\url{#2}}

\bibitem[{\citenamefont{Strocchi}(2005, 2008)}]{Strocchi2005}
\bibinfo{author}{\bibfnamefont{F.}~\bibnamefont{Strocchi}},
  \emph{\bibinfo{title}{Symmetry Breaking}}, The Lecture Notes in Physics
  (\bibinfo{publisher}{Springer-Verlag}, \bibinfo{address}{{Berlin
  Heidelberg}}, \bibinfo{year}{2005, 2008}), ISBN
  \bibinfo{isbn}{978-3-540-73592-2}.

\bibitem[{\citenamefont{Wilczek}(2012)}]{Wilczek2012}
\bibinfo{author}{\bibfnamefont{F.}~\bibnamefont{Wilczek}},
  \bibinfo{journal}{Phys. Rev. Lett.} \textbf{\bibinfo{volume}{109}},
  \bibinfo{pages}{160401} (\bibinfo{year}{2012}),
  \urlprefix\url{http://link.aps.org/doi/10.1103/PhysRevLett.109.160401}.

\bibitem[{\citenamefont{{Sacha} and {Zakrzewski}}(2018)}]{Sacha2017rev}
\bibinfo{author}{\bibfnamefont{K.}~\bibnamefont{{Sacha}}} \bibnamefont{and}
  \bibinfo{author}{\bibfnamefont{J.}~\bibnamefont{{Zakrzewski}}},
  \bibinfo{journal}{Rep. Prog. Phys.} \textbf{\bibinfo{volume}{81}},
  \bibinfo{pages}{016401} (\bibinfo{year}{2018}),
  \urlprefix\url{https://doi.org/10.1088/1361-6633/aa8b38}.

\bibitem[{\citenamefont{Bruno}(2013)}]{Bruno2013b}
\bibinfo{author}{\bibfnamefont{P.}~\bibnamefont{Bruno}},
  \bibinfo{journal}{Phys. Rev. Lett.} \textbf{\bibinfo{volume}{111}},
  \bibinfo{pages}{070402} (\bibinfo{year}{2013}),
  \urlprefix\url{http://link.aps.org/doi/10.1103/PhysRevLett.111.070402}.

\bibitem[{\citenamefont{Watanabe and Oshikawa}(2015)}]{Watanabe2015}
\bibinfo{author}{\bibfnamefont{H.}~\bibnamefont{Watanabe}} \bibnamefont{and}
  \bibinfo{author}{\bibfnamefont{M.}~\bibnamefont{Oshikawa}},
  \bibinfo{journal}{Phys. Rev. Lett.} \textbf{\bibinfo{volume}{114}},
  \bibinfo{pages}{251603} (\bibinfo{year}{2015}),
  \urlprefix\url{http://link.aps.org/doi/10.1103/PhysRevLett.114.251603}.

\bibitem[{\citenamefont{Syrwid et~al.}(2017)\citenamefont{Syrwid, Zakrzewski,
  and Sacha}}]{Syrwid2017}
\bibinfo{author}{\bibfnamefont{A.}~\bibnamefont{Syrwid}},
  \bibinfo{author}{\bibfnamefont{J.}~\bibnamefont{Zakrzewski}},
  \bibnamefont{and} \bibinfo{author}{\bibfnamefont{K.}~\bibnamefont{Sacha}},
  \bibinfo{journal}{Phys. Rev. Lett.} \textbf{\bibinfo{volume}{119}},
  \bibinfo{pages}{250602} (\bibinfo{year}{2017}),
  \urlprefix\url{https://link.aps.org/doi/10.1103/PhysRevLett.119.250602}.

\bibitem[{\citenamefont{Shapere and Wilczek}(2012)}]{Shapere2012}
\bibinfo{author}{\bibfnamefont{A.}~\bibnamefont{Shapere}} \bibnamefont{and}
  \bibinfo{author}{\bibfnamefont{F.}~\bibnamefont{Wilczek}},
  \bibinfo{journal}{Phys. Rev. Lett.} \textbf{\bibinfo{volume}{109}},
  \bibinfo{pages}{160402} (\bibinfo{year}{2012}),
  \urlprefix\url{http://link.aps.org/doi/10.1103/PhysRevLett.109.160402}.

\bibitem[{\citenamefont{Ghosh}(2014)}]{Ghosh2014}
\bibinfo{author}{\bibfnamefont{S.}~\bibnamefont{Ghosh}},
  \bibinfo{journal}{Physica A: Statistical Mechanics and its Applications}
  \textbf{\bibinfo{volume}{407}}, \bibinfo{pages}{245 } (\bibinfo{year}{2014}),
  ISSN \bibinfo{issn}{0378-4371},
  \urlprefix\url{http://www.sciencedirect.com/science/article/pii/S0378437114003161}.

\bibitem[{\citenamefont{{Yao} et~al.}(2018)\citenamefont{{Yao}, {Nayak},
  {Balents}, and {Zaletel}}}]{Yao2018}
\bibinfo{author}{\bibfnamefont{N.~Y.} \bibnamefont{{Yao}}},
  \bibinfo{author}{\bibfnamefont{C.}~\bibnamefont{{Nayak}}},
  \bibinfo{author}{\bibfnamefont{L.}~\bibnamefont{{Balents}}},
  \bibnamefont{and} \bibinfo{author}{\bibfnamefont{M.~P.}
  \bibnamefont{{Zaletel}}}, \bibinfo{journal}{ArXiv e-prints}
  (\bibinfo{year}{2018}), \eprint{1801.02628}.

\bibitem[{\citenamefont{Das et~al.}(2018)\citenamefont{Das, Pan, Ghosh, and
  Pal}}]{Das2018}
\bibinfo{author}{\bibfnamefont{P.}~\bibnamefont{Das}},
  \bibinfo{author}{\bibfnamefont{S.}~\bibnamefont{Pan}},
  \bibinfo{author}{\bibfnamefont{S.}~\bibnamefont{Ghosh}}, \bibnamefont{and}
  \bibinfo{author}{\bibfnamefont{P.}~\bibnamefont{Pal}},
  \bibinfo{journal}{Phys. Rev. D} \textbf{\bibinfo{volume}{98}},
  \bibinfo{pages}{024004} (\bibinfo{year}{2018}),
  \urlprefix\url{https://link.aps.org/doi/10.1103/PhysRevD.98.024004}.

\bibitem[{\citenamefont{Alvarez et~al.}(2017)\citenamefont{Alvarez, Canfora,
  Dimakis, and Paliathanasis}}]{Alvarez2017}
\bibinfo{author}{\bibfnamefont{P.}~\bibnamefont{Alvarez}},
  \bibinfo{author}{\bibfnamefont{F.}~\bibnamefont{Canfora}},
  \bibinfo{author}{\bibfnamefont{N.}~\bibnamefont{Dimakis}}, \bibnamefont{and}
  \bibinfo{author}{\bibfnamefont{A.}~\bibnamefont{Paliathanasis}},
  \bibinfo{journal}{Physics Letters B} \textbf{\bibinfo{volume}{773}},
  \bibinfo{pages}{401 } (\bibinfo{year}{2017}), ISSN \bibinfo{issn}{0370-2693},
  \urlprefix\url{http://www.sciencedirect.com/science/article/pii/S0370269317306950}.

\bibitem[{\citenamefont{Avil\'es et~al.}(2017)\citenamefont{Avil\'es, Canfora,
  Dimakis, and Hidalgo}}]{Aviles2017}
\bibinfo{author}{\bibfnamefont{L.}~\bibnamefont{Avil\'es}},
  \bibinfo{author}{\bibfnamefont{F.}~\bibnamefont{Canfora}},
  \bibinfo{author}{\bibfnamefont{N.}~\bibnamefont{Dimakis}}, \bibnamefont{and}
  \bibinfo{author}{\bibfnamefont{D.}~\bibnamefont{Hidalgo}},
  \bibinfo{journal}{Phys. Rev. D} \textbf{\bibinfo{volume}{96}},
  \bibinfo{pages}{125005} (\bibinfo{year}{2017}),
  \urlprefix\url{https://link.aps.org/doi/10.1103/PhysRevD.96.125005}.

\bibitem[{\citenamefont{Flicker}(2018)}]{Flicker2018}
\bibinfo{author}{\bibfnamefont{F.}~\bibnamefont{Flicker}},
  \bibinfo{journal}{SciPost Phys.} \textbf{\bibinfo{volume}{5}},
  \bibinfo{pages}{1} (\bibinfo{year}{2018}),
  \urlprefix\url{https://scipost.org/10.21468/SciPostPhys.5.1.001}.

\bibitem[{\citenamefont{Sacha}(2015{\natexlab{a}})}]{Sacha2015}
\bibinfo{author}{\bibfnamefont{K.}~\bibnamefont{Sacha}},
  \bibinfo{journal}{Phys. Rev. A} \textbf{\bibinfo{volume}{91}},
  \bibinfo{pages}{033617} (\bibinfo{year}{2015}{\natexlab{a}}),
  \urlprefix\url{http://link.aps.org/doi/10.1103/PhysRevA.91.033617}.

\bibitem[{\citenamefont{Khemani et~al.}(2016)\citenamefont{Khemani, Lazarides,
  Moessner, and Sondhi}}]{Khemani16}
\bibinfo{author}{\bibfnamefont{V.}~\bibnamefont{Khemani}},
  \bibinfo{author}{\bibfnamefont{A.}~\bibnamefont{Lazarides}},
  \bibinfo{author}{\bibfnamefont{R.}~\bibnamefont{Moessner}}, \bibnamefont{and}
  \bibinfo{author}{\bibfnamefont{S.~L.} \bibnamefont{Sondhi}},
  \bibinfo{journal}{Phys. Rev. Lett.} \textbf{\bibinfo{volume}{116}},
  \bibinfo{pages}{250401} (\bibinfo{year}{2016}),
  \urlprefix\url{http://link.aps.org/doi/10.1103/PhysRevLett.116.250401}.

\bibitem[{\citenamefont{Else et~al.}(2016)\citenamefont{Else, Bauer, and
  Nayak}}]{ElseFTC}
\bibinfo{author}{\bibfnamefont{D.~V.} \bibnamefont{Else}},
  \bibinfo{author}{\bibfnamefont{B.}~\bibnamefont{Bauer}}, \bibnamefont{and}
  \bibinfo{author}{\bibfnamefont{C.}~\bibnamefont{Nayak}},
  \bibinfo{journal}{Phys. Rev. Lett.} \textbf{\bibinfo{volume}{117}},
  \bibinfo{pages}{090402} (\bibinfo{year}{2016}),
  \urlprefix\url{http://link.aps.org/doi/10.1103/PhysRevLett.117.090402}.

\bibitem[{\citenamefont{Yao et~al.}(2017)\citenamefont{Yao, Potter, Potirniche,
  and Vishwanath}}]{Yao2017}
\bibinfo{author}{\bibfnamefont{N.~Y.} \bibnamefont{Yao}},
  \bibinfo{author}{\bibfnamefont{A.~C.} \bibnamefont{Potter}},
  \bibinfo{author}{\bibfnamefont{I.-D.} \bibnamefont{Potirniche}},
  \bibnamefont{and}
  \bibinfo{author}{\bibfnamefont{A.}~\bibnamefont{Vishwanath}},
  \bibinfo{journal}{Phys. Rev. Lett.} \textbf{\bibinfo{volume}{118}},
  \bibinfo{pages}{030401} (\bibinfo{year}{2017}),
  \urlprefix\url{http://link.aps.org/doi/10.1103/PhysRevLett.118.030401}.

\bibitem[{\citenamefont{Zhang et~al.}(2017)\citenamefont{Zhang, Hess,
  Kyprianidis, Becker, Lee, Smith, Pagano, Potirniche, Potter, Vishwanath
  et~al.}}]{Zhang2017}
\bibinfo{author}{\bibfnamefont{J.}~\bibnamefont{Zhang}},
  \bibinfo{author}{\bibfnamefont{P.~W.} \bibnamefont{Hess}},
  \bibinfo{author}{\bibfnamefont{A.}~\bibnamefont{Kyprianidis}},
  \bibinfo{author}{\bibfnamefont{P.}~\bibnamefont{Becker}},
  \bibinfo{author}{\bibfnamefont{A.}~\bibnamefont{Lee}},
  \bibinfo{author}{\bibfnamefont{J.}~\bibnamefont{Smith}},
  \bibinfo{author}{\bibfnamefont{G.}~\bibnamefont{Pagano}},
  \bibinfo{author}{\bibfnamefont{I.-D.} \bibnamefont{Potirniche}},
  \bibinfo{author}{\bibfnamefont{A.~C.} \bibnamefont{Potter}},
  \bibinfo{author}{\bibfnamefont{A.}~\bibnamefont{Vishwanath}},
  \bibnamefont{et~al.}, \bibinfo{journal}{Nature}
  \textbf{\bibinfo{volume}{543}}, \bibinfo{pages}{217} (\bibinfo{year}{2017}),
  ISSN \bibinfo{issn}{0028-0836}, \bibinfo{note}{letter},
  \urlprefix\url{http://dx.doi.org/10.1038/nature21413}.

\bibitem[{\citenamefont{Choi et~al.}(2017)\citenamefont{Choi, Choi, Landig,
  Kucsko, Zhou, Isoya, Jelezko, Onoda, Sumiya, Khemani et~al.}}]{Choi2017}
\bibinfo{author}{\bibfnamefont{S.}~\bibnamefont{Choi}},
  \bibinfo{author}{\bibfnamefont{J.}~\bibnamefont{Choi}},
  \bibinfo{author}{\bibfnamefont{R.}~\bibnamefont{Landig}},
  \bibinfo{author}{\bibfnamefont{G.}~\bibnamefont{Kucsko}},
  \bibinfo{author}{\bibfnamefont{H.}~\bibnamefont{Zhou}},
  \bibinfo{author}{\bibfnamefont{J.}~\bibnamefont{Isoya}},
  \bibinfo{author}{\bibfnamefont{F.}~\bibnamefont{Jelezko}},
  \bibinfo{author}{\bibfnamefont{S.}~\bibnamefont{Onoda}},
  \bibinfo{author}{\bibfnamefont{H.}~\bibnamefont{Sumiya}},
  \bibinfo{author}{\bibfnamefont{V.}~\bibnamefont{Khemani}},
  \bibnamefont{et~al.}, \bibinfo{journal}{Nature}
  \textbf{\bibinfo{volume}{543}}, \bibinfo{pages}{221} (\bibinfo{year}{2017}),
  ISSN \bibinfo{issn}{0028-0836}, \bibinfo{note}{letter},
  \urlprefix\url{http://dx.doi.org/10.1038/nature21426}.

\bibitem[{\citenamefont{Shirley}(1965)}]{Shirley1965}
\bibinfo{author}{\bibfnamefont{J.~H.} \bibnamefont{Shirley}},
  \bibinfo{journal}{Phys. Rev.} \textbf{\bibinfo{volume}{138}},
  \bibinfo{pages}{B979} (\bibinfo{year}{1965}),
  \urlprefix\url{https://link.aps.org/doi/10.1103/PhysRev.138.B979}.

\bibitem[{\citenamefont{{Giergiel} et~al.}(2018)\citenamefont{{Giergiel},
  {Kuro{\'s}}, and {Sacha}}}]{Giergiel2018c}
\bibinfo{author}{\bibfnamefont{K.}~\bibnamefont{{Giergiel}}},
  \bibinfo{author}{\bibfnamefont{A.}~\bibnamefont{{Kuro{\'s}}}},
  \bibnamefont{and} \bibinfo{author}{\bibfnamefont{K.}~\bibnamefont{{Sacha}}},
  \bibinfo{journal}{ArXiv e-prints}  (\bibinfo{year}{2018}),
  \eprint{1807.02105}.

\bibitem[{\citenamefont{Sacha}(2015{\natexlab{b}})}]{sacha15a}
\bibinfo{author}{\bibfnamefont{K.}~\bibnamefont{Sacha}}, \bibinfo{journal}{Sci.
  Rep.} \textbf{\bibinfo{volume}{5}}, \bibinfo{pages}{10787}
  (\bibinfo{year}{2015}{\natexlab{b}}),
  \urlprefix\url{https://www.nature.com/articles/srep10787}.

\bibitem[{\citenamefont{Delande et~al.}(2017)\citenamefont{Delande,
  Morales-Molina, and Sacha}}]{delande17}
\bibinfo{author}{\bibfnamefont{D.}~\bibnamefont{Delande}},
  \bibinfo{author}{\bibfnamefont{L.}~\bibnamefont{Morales-Molina}},
  \bibnamefont{and} \bibinfo{author}{\bibfnamefont{K.}~\bibnamefont{Sacha}},
  \bibinfo{journal}{Phys. Rev. Lett.} \textbf{\bibinfo{volume}{119}},
  \bibinfo{pages}{230404} (\bibinfo{year}{2017}),
  \urlprefix\url{https://link.aps.org/doi/10.1103/PhysRevLett.119.230404}.

\bibitem[{\citenamefont{Giergiel and Sacha}(2017)}]{Giergiel2017}
\bibinfo{author}{\bibfnamefont{K.}~\bibnamefont{Giergiel}} \bibnamefont{and}
  \bibinfo{author}{\bibfnamefont{K.}~\bibnamefont{Sacha}},
  \bibinfo{journal}{Phys. Rev. A} \textbf{\bibinfo{volume}{95}},
  \bibinfo{pages}{063402} (\bibinfo{year}{2017}),
  \urlprefix\url{https://link.aps.org/doi/10.1103/PhysRevA.95.063402}.

\bibitem[{\citenamefont{Lazarides and Moessner}(2017)}]{Lazarides2017}
\bibinfo{author}{\bibfnamefont{A.}~\bibnamefont{Lazarides}} \bibnamefont{and}
  \bibinfo{author}{\bibfnamefont{R.}~\bibnamefont{Moessner}},
  \bibinfo{journal}{Phys. Rev. B} \textbf{\bibinfo{volume}{95}},
  \bibinfo{pages}{195135} (\bibinfo{year}{2017}),
  \urlprefix\url{https://link.aps.org/doi/10.1103/PhysRevB.95.195135}.

\bibitem[{\citenamefont{Russomanno et~al.}(2017)\citenamefont{Russomanno,
  Iemini, Dalmonte, and Fazio}}]{Russomanno2017}
\bibinfo{author}{\bibfnamefont{A.}~\bibnamefont{Russomanno}},
  \bibinfo{author}{\bibfnamefont{F.}~\bibnamefont{Iemini}},
  \bibinfo{author}{\bibfnamefont{M.}~\bibnamefont{Dalmonte}}, \bibnamefont{and}
  \bibinfo{author}{\bibfnamefont{R.}~\bibnamefont{Fazio}},
  \bibinfo{journal}{Phys. Rev. B} \textbf{\bibinfo{volume}{95}},
  \bibinfo{pages}{214307} (\bibinfo{year}{2017}),
  \urlprefix\url{https://link.aps.org/doi/10.1103/PhysRevB.95.214307}.

\bibitem[{\citenamefont{Zeng and Sheng}(2017)}]{Zeng2017}
\bibinfo{author}{\bibfnamefont{T.-S.} \bibnamefont{Zeng}} \bibnamefont{and}
  \bibinfo{author}{\bibfnamefont{D.~N.} \bibnamefont{Sheng}},
  \bibinfo{journal}{Phys. Rev. B} \textbf{\bibinfo{volume}{96}},
  \bibinfo{pages}{094202} (\bibinfo{year}{2017}),
  \urlprefix\url{https://link.aps.org/doi/10.1103/PhysRevB.96.094202}.

\bibitem[{\citenamefont{Nakatsugawa et~al.}(2017)\citenamefont{Nakatsugawa,
  Fujii, and Tanda}}]{Nakatsugawa2017}
\bibinfo{author}{\bibfnamefont{K.}~\bibnamefont{Nakatsugawa}},
  \bibinfo{author}{\bibfnamefont{T.}~\bibnamefont{Fujii}}, \bibnamefont{and}
  \bibinfo{author}{\bibfnamefont{S.}~\bibnamefont{Tanda}},
  \bibinfo{journal}{Phys. Rev. B} \textbf{\bibinfo{volume}{96}},
  \bibinfo{pages}{094308} (\bibinfo{year}{2017}),
  \urlprefix\url{https://link.aps.org/doi/10.1103/PhysRevB.96.094308}.

\bibitem[{\citenamefont{Ho et~al.}(2017)\citenamefont{Ho, Choi, Lukin, and
  Abanin}}]{Ho2017}
\bibinfo{author}{\bibfnamefont{W.~W.} \bibnamefont{Ho}},
  \bibinfo{author}{\bibfnamefont{S.}~\bibnamefont{Choi}},
  \bibinfo{author}{\bibfnamefont{M.~D.} \bibnamefont{Lukin}}, \bibnamefont{and}
  \bibinfo{author}{\bibfnamefont{D.~A.} \bibnamefont{Abanin}},
  \bibinfo{journal}{Phys. Rev. Lett.} \textbf{\bibinfo{volume}{119}},
  \bibinfo{pages}{010602} (\bibinfo{year}{2017}),
  \urlprefix\url{https://link.aps.org/doi/10.1103/PhysRevLett.119.010602}.

\bibitem[{\citenamefont{Huang et~al.}(2018{\natexlab{a}})\citenamefont{Huang,
  Wu, and Liu}}]{Huang2017}
\bibinfo{author}{\bibfnamefont{B.}~\bibnamefont{Huang}},
  \bibinfo{author}{\bibfnamefont{Y.-H.} \bibnamefont{Wu}}, \bibnamefont{and}
  \bibinfo{author}{\bibfnamefont{W.~V.} \bibnamefont{Liu}},
  \bibinfo{journal}{Phys. Rev. Lett.} \textbf{\bibinfo{volume}{120}},
  \bibinfo{pages}{110603} (\bibinfo{year}{2018}{\natexlab{a}}),
  \urlprefix\url{https://link.aps.org/doi/10.1103/PhysRevLett.120.110603}.

\bibitem[{\citenamefont{Gong et~al.}(2018)\citenamefont{Gong, Hamazaki, and
  Ueda}}]{Gong2017}
\bibinfo{author}{\bibfnamefont{Z.}~\bibnamefont{Gong}},
  \bibinfo{author}{\bibfnamefont{R.}~\bibnamefont{Hamazaki}}, \bibnamefont{and}
  \bibinfo{author}{\bibfnamefont{M.}~\bibnamefont{Ueda}},
  \bibinfo{journal}{Phys. Rev. Lett.} \textbf{\bibinfo{volume}{120}},
  \bibinfo{pages}{040404} (\bibinfo{year}{2018}),
  \urlprefix\url{https://link.aps.org/doi/10.1103/PhysRevLett.120.040404}.

\bibitem[{\citenamefont{Wang et~al.}(2018)\citenamefont{Wang, Xing, Carlo, and
  Poletti}}]{Wang2017}
\bibinfo{author}{\bibfnamefont{R.~R.~W.} \bibnamefont{Wang}},
  \bibinfo{author}{\bibfnamefont{B.}~\bibnamefont{Xing}},
  \bibinfo{author}{\bibfnamefont{G.~G.} \bibnamefont{Carlo}}, \bibnamefont{and}
  \bibinfo{author}{\bibfnamefont{D.}~\bibnamefont{Poletti}},
  \bibinfo{journal}{Phys. Rev. E} \textbf{\bibinfo{volume}{97}},
  \bibinfo{pages}{020202} (\bibinfo{year}{2018}),
  \urlprefix\url{https://link.aps.org/doi/10.1103/PhysRevE.97.020202}.

\bibitem[{\citenamefont{Iemini et~al.}(2018)\citenamefont{Iemini, Russomanno,
  Keeling, Schir\`o, Dalmonte, and Fazio}}]{Iemini2017}
\bibinfo{author}{\bibfnamefont{F.}~\bibnamefont{Iemini}},
  \bibinfo{author}{\bibfnamefont{A.}~\bibnamefont{Russomanno}},
  \bibinfo{author}{\bibfnamefont{J.}~\bibnamefont{Keeling}},
  \bibinfo{author}{\bibfnamefont{M.}~\bibnamefont{Schir\`o}},
  \bibinfo{author}{\bibfnamefont{M.}~\bibnamefont{Dalmonte}}, \bibnamefont{and}
  \bibinfo{author}{\bibfnamefont{R.}~\bibnamefont{Fazio}},
  \bibinfo{journal}{Phys. Rev. Lett.} \textbf{\bibinfo{volume}{121}},
  \bibinfo{pages}{035301} (\bibinfo{year}{2018}),
  \urlprefix\url{https://link.aps.org/doi/10.1103/PhysRevLett.121.035301}.

\bibitem[{\citenamefont{Mierzejewski et~al.}(2017)\citenamefont{Mierzejewski,
  Giergiel, and Sacha}}]{Mierzejewski2017}
\bibinfo{author}{\bibfnamefont{M.}~\bibnamefont{Mierzejewski}},
  \bibinfo{author}{\bibfnamefont{K.}~\bibnamefont{Giergiel}}, \bibnamefont{and}
  \bibinfo{author}{\bibfnamefont{K.}~\bibnamefont{Sacha}},
  \bibinfo{journal}{Phys. Rev. B} \textbf{\bibinfo{volume}{96}},
  \bibinfo{pages}{140201} (\bibinfo{year}{2017}),
  \urlprefix\url{https://link.aps.org/doi/10.1103/PhysRevB.96.140201}.

\bibitem[{\citenamefont{{Prokof'ev} and {Svistunov}}(2017)}]{Prokofev2017}
\bibinfo{author}{\bibfnamefont{N.~V.} \bibnamefont{{Prokof'ev}}}
  \bibnamefont{and} \bibinfo{author}{\bibfnamefont{B.~V.}
  \bibnamefont{{Svistunov}}}, \bibinfo{journal}{ArXiv e-prints}
  (\bibinfo{year}{2017}), \eprint{1710.00721}.

\bibitem[{\citenamefont{Pal et~al.}(2018)\citenamefont{Pal, Nishad, Mahesh, and
  Sreejith}}]{Pal2018}
\bibinfo{author}{\bibfnamefont{S.}~\bibnamefont{Pal}},
  \bibinfo{author}{\bibfnamefont{N.}~\bibnamefont{Nishad}},
  \bibinfo{author}{\bibfnamefont{T.~S.} \bibnamefont{Mahesh}},
  \bibnamefont{and} \bibinfo{author}{\bibfnamefont{G.~J.}
  \bibnamefont{Sreejith}}, \bibinfo{journal}{Phys. Rev. Lett.}
  \textbf{\bibinfo{volume}{120}}, \bibinfo{pages}{180602}
  (\bibinfo{year}{2018}),
  \urlprefix\url{https://link.aps.org/doi/10.1103/PhysRevLett.120.180602}.

\bibitem[{\citenamefont{Rovny et~al.}(2018{\natexlab{a}})\citenamefont{Rovny,
  Blum, and Barrett}}]{Rovny2018}
\bibinfo{author}{\bibfnamefont{J.}~\bibnamefont{Rovny}},
  \bibinfo{author}{\bibfnamefont{R.~L.} \bibnamefont{Blum}}, \bibnamefont{and}
  \bibinfo{author}{\bibfnamefont{S.~E.} \bibnamefont{Barrett}},
  \bibinfo{journal}{Phys. Rev. Lett.} \textbf{\bibinfo{volume}{120}},
  \bibinfo{pages}{180603} (\bibinfo{year}{2018}{\natexlab{a}}),
  \urlprefix\url{https://link.aps.org/doi/10.1103/PhysRevLett.120.180603}.

\bibitem[{\citenamefont{Rovny et~al.}(2018{\natexlab{b}})\citenamefont{Rovny,
  Blum, and Barrett}}]{Rovny2018a}
\bibinfo{author}{\bibfnamefont{J.}~\bibnamefont{Rovny}},
  \bibinfo{author}{\bibfnamefont{R.~L.} \bibnamefont{Blum}}, \bibnamefont{and}
  \bibinfo{author}{\bibfnamefont{S.~E.} \bibnamefont{Barrett}},
  \bibinfo{journal}{Phys. Rev. B} \textbf{\bibinfo{volume}{97}},
  \bibinfo{pages}{184301} (\bibinfo{year}{2018}{\natexlab{b}}),
  \urlprefix\url{https://link.aps.org/doi/10.1103/PhysRevB.97.184301}.

\bibitem[{\citenamefont{Giergiel
  et~al.}(2018{\natexlab{a}})\citenamefont{Giergiel, Miroszewski, and
  Sacha}}]{Giergiel2018}
\bibinfo{author}{\bibfnamefont{K.}~\bibnamefont{Giergiel}},
  \bibinfo{author}{\bibfnamefont{A.}~\bibnamefont{Miroszewski}},
  \bibnamefont{and} \bibinfo{author}{\bibfnamefont{K.}~\bibnamefont{Sacha}},
  \bibinfo{journal}{Phys. Rev. Lett.} \textbf{\bibinfo{volume}{120}},
  \bibinfo{pages}{140401} (\bibinfo{year}{2018}{\natexlab{a}}),
  \urlprefix\url{https://link.aps.org/doi/10.1103/PhysRevLett.120.140401}.

\bibitem[{\citenamefont{Bomantara and Gong}(2018)}]{Bomantara2018}
\bibinfo{author}{\bibfnamefont{R.~W.} \bibnamefont{Bomantara}}
  \bibnamefont{and} \bibinfo{author}{\bibfnamefont{J.}~\bibnamefont{Gong}},
  \bibinfo{journal}{Phys. Rev. Lett.} \textbf{\bibinfo{volume}{120}},
  \bibinfo{pages}{230405} (\bibinfo{year}{2018}),
  \urlprefix\url{https://link.aps.org/doi/10.1103/PhysRevLett.120.230405}.

\bibitem[{\citenamefont{Kosior and Sacha}(2018)}]{Kosior2018}
\bibinfo{author}{\bibfnamefont{A.}~\bibnamefont{Kosior}} \bibnamefont{and}
  \bibinfo{author}{\bibfnamefont{K.}~\bibnamefont{Sacha}},
  \bibinfo{journal}{Phys. Rev. A} \textbf{\bibinfo{volume}{97}},
  \bibinfo{pages}{053621} (\bibinfo{year}{2018}),
  \urlprefix\url{https://link.aps.org/doi/10.1103/PhysRevA.97.053621}.

\bibitem[{\citenamefont{Mizuta et~al.}(2018)\citenamefont{Mizuta, Takasan,
  Nakagawa, and Kawakami}}]{Mizuta2018}
\bibinfo{author}{\bibfnamefont{K.}~\bibnamefont{Mizuta}},
  \bibinfo{author}{\bibfnamefont{K.}~\bibnamefont{Takasan}},
  \bibinfo{author}{\bibfnamefont{M.}~\bibnamefont{Nakagawa}}, \bibnamefont{and}
  \bibinfo{author}{\bibfnamefont{N.}~\bibnamefont{Kawakami}},
  \bibinfo{journal}{Phys. Rev. Lett.} \textbf{\bibinfo{volume}{121}},
  \bibinfo{pages}{093001} (\bibinfo{year}{2018}),
  \urlprefix\url{https://link.aps.org/doi/10.1103/PhysRevLett.121.093001}.

\bibitem[{\citenamefont{Giergiel
  et~al.}(2018{\natexlab{b}})\citenamefont{Giergiel, Kosior, Hannaford, and
  Sacha}}]{Giergiel2018a}
\bibinfo{author}{\bibfnamefont{K.}~\bibnamefont{Giergiel}},
  \bibinfo{author}{\bibfnamefont{A.}~\bibnamefont{Kosior}},
  \bibinfo{author}{\bibfnamefont{P.}~\bibnamefont{Hannaford}},
  \bibnamefont{and} \bibinfo{author}{\bibfnamefont{K.}~\bibnamefont{Sacha}},
  \bibinfo{journal}{Phys. Rev. A} \textbf{\bibinfo{volume}{98}},
  \bibinfo{pages}{013613} (\bibinfo{year}{2018}{\natexlab{b}}),
  \urlprefix\url{https://link.aps.org/doi/10.1103/PhysRevA.98.013613}.

\bibitem[{\citenamefont{{Giergiel} et~al.}(2018)\citenamefont{{Giergiel},
  {Dauphin}, {Lewenstein}, {Zakrzewski}, and {Sacha}}}]{Giergiel2018b}
\bibinfo{author}{\bibfnamefont{K.}~\bibnamefont{{Giergiel}}},
  \bibinfo{author}{\bibfnamefont{A.}~\bibnamefont{{Dauphin}}},
  \bibinfo{author}{\bibfnamefont{M.}~\bibnamefont{{Lewenstein}}},
  \bibinfo{author}{\bibfnamefont{J.}~\bibnamefont{{Zakrzewski}}},
  \bibnamefont{and} \bibinfo{author}{\bibfnamefont{K.}~\bibnamefont{{Sacha}}},
  \bibinfo{journal}{ArXiv e-prints}  (\bibinfo{year}{2018}),
  \eprint{1806.10536}.

\bibitem[{\citenamefont{Autti et~al.}(2018)\citenamefont{Autti, Eltsov, and
  Volovik}}]{Autti2018}
\bibinfo{author}{\bibfnamefont{S.}~\bibnamefont{Autti}},
  \bibinfo{author}{\bibfnamefont{V.~B.} \bibnamefont{Eltsov}},
  \bibnamefont{and} \bibinfo{author}{\bibfnamefont{G.~E.}
  \bibnamefont{Volovik}}, \bibinfo{journal}{Phys. Rev. Lett.}
  \textbf{\bibinfo{volume}{120}}, \bibinfo{pages}{215301}
  (\bibinfo{year}{2018}),
  \urlprefix\url{https://link.aps.org/doi/10.1103/PhysRevLett.120.215301}.

\bibitem[{\citenamefont{{Tucker} et~al.}(2018)\citenamefont{{Tucker}, {Zhu},
  {Lewis-Swan}, {Marino}, {Jimenez}, {Restrepo}, and {Rey}}}]{Tucker2018}
\bibinfo{author}{\bibfnamefont{K.}~\bibnamefont{{Tucker}}},
  \bibinfo{author}{\bibfnamefont{B.}~\bibnamefont{{Zhu}}},
  \bibinfo{author}{\bibfnamefont{R.~J.} \bibnamefont{{Lewis-Swan}}},
  \bibinfo{author}{\bibfnamefont{J.}~\bibnamefont{{Marino}}},
  \bibinfo{author}{\bibfnamefont{F.}~\bibnamefont{{Jimenez}}},
  \bibinfo{author}{\bibfnamefont{J.~G.} \bibnamefont{{Restrepo}}},
  \bibnamefont{and} \bibinfo{author}{\bibfnamefont{A.~M.} \bibnamefont{{Rey}}},
  \bibinfo{journal}{ArXiv e-prints}  (\bibinfo{year}{2018}),
  \eprint{1805.03343}.

\bibitem[{\citenamefont{Kosior et~al.}(2018)\citenamefont{Kosior, Syrwid, and
  Sacha}}]{Kosior2018a}
\bibinfo{author}{\bibfnamefont{A.}~\bibnamefont{Kosior}},
  \bibinfo{author}{\bibfnamefont{A.}~\bibnamefont{Syrwid}}, \bibnamefont{and}
  \bibinfo{author}{\bibfnamefont{K.}~\bibnamefont{Sacha}},
  \bibinfo{journal}{Phys. Rev. A} \textbf{\bibinfo{volume}{98}},
  \bibinfo{pages}{023612} (\bibinfo{year}{2018}),
  \urlprefix\url{https://link.aps.org/doi/10.1103/PhysRevA.98.023612}.

\bibitem[{\citenamefont{{Yu} et~al.}(2018)\citenamefont{{Yu}, {Tangpanitanon},
  {Glaetzle}, {Jaksch}, and {Angelakis}}}]{Yu2018}
\bibinfo{author}{\bibfnamefont{W.~C.} \bibnamefont{{Yu}}},
  \bibinfo{author}{\bibfnamefont{J.}~\bibnamefont{{Tangpanitanon}}},
  \bibinfo{author}{\bibfnamefont{A.~W.} \bibnamefont{{Glaetzle}}},
  \bibinfo{author}{\bibfnamefont{D.}~\bibnamefont{{Jaksch}}}, \bibnamefont{and}
  \bibinfo{author}{\bibfnamefont{D.~G.} \bibnamefont{{Angelakis}}},
  \bibinfo{journal}{ArXiv e-prints}  (\bibinfo{year}{2018}),
  \eprint{1807.07738}.

\bibitem[{\citenamefont{{O'Sullivan} et~al.}(2018)\citenamefont{{O'Sullivan},
  {Lunt}, {Zollitsch}, {Thewalt}, {Morton}, and {Pal}}}]{Sullivan2018}
\bibinfo{author}{\bibfnamefont{J.}~\bibnamefont{{O'Sullivan}}},
  \bibinfo{author}{\bibfnamefont{O.}~\bibnamefont{{Lunt}}},
  \bibinfo{author}{\bibfnamefont{C.~W.} \bibnamefont{{Zollitsch}}},
  \bibinfo{author}{\bibfnamefont{M.~L.~W.} \bibnamefont{{Thewalt}}},
  \bibinfo{author}{\bibfnamefont{J.~J.~L.} \bibnamefont{{Morton}}},
  \bibnamefont{and} \bibinfo{author}{\bibfnamefont{A.}~\bibnamefont{{Pal}}},
  \bibinfo{journal}{ArXiv e-prints}  (\bibinfo{year}{2018}),
  \eprint{1807.09884}.

\bibitem[{\citenamefont{Smits et~al.}(2018)\citenamefont{Smits, Liao, Stoof,
  and van~der Straten}}]{Smits2018}
\bibinfo{author}{\bibfnamefont{J.}~\bibnamefont{Smits}},
  \bibinfo{author}{\bibfnamefont{L.}~\bibnamefont{Liao}},
  \bibinfo{author}{\bibfnamefont{H.~T.~C.} \bibnamefont{Stoof}},
  \bibnamefont{and} \bibinfo{author}{\bibfnamefont{P.}~\bibnamefont{van~der
  Straten}}, \bibinfo{journal}{Phys. Rev. Lett.}
  \textbf{\bibinfo{volume}{121}}, \bibinfo{pages}{185301}
  (\bibinfo{year}{2018}),
  \urlprefix\url{https://link.aps.org/doi/10.1103/PhysRevLett.121.185301}.

\bibitem[{\citenamefont{{Gambetta} et~al.}(2018)\citenamefont{{Gambetta},
  {Carollo}, {Marcuzzi}, {Garrahan}, and {Lesanovsky}}}]{Gambetta2018}
\bibinfo{author}{\bibfnamefont{F.~M.} \bibnamefont{{Gambetta}}},
  \bibinfo{author}{\bibfnamefont{F.}~\bibnamefont{{Carollo}}},
  \bibinfo{author}{\bibfnamefont{M.}~\bibnamefont{{Marcuzzi}}},
  \bibinfo{author}{\bibfnamefont{J.~P.} \bibnamefont{{Garrahan}}},
  \bibnamefont{and}
  \bibinfo{author}{\bibfnamefont{I.}~\bibnamefont{{Lesanovsky}}},
  \bibinfo{journal}{arXiv e-prints} \bibinfo{eid}{arXiv:1807.10161}
  (\bibinfo{year}{2018}), \eprint{1807.10161}.

\bibitem[{\citenamefont{{Liao} et~al.}(2018)\citenamefont{{Liao}, {Smits}, {van
  der Straten}, and {Stoof}}}]{Liao2018}
\bibinfo{author}{\bibfnamefont{L.}~\bibnamefont{{Liao}}},
  \bibinfo{author}{\bibfnamefont{J.}~\bibnamefont{{Smits}}},
  \bibinfo{author}{\bibfnamefont{P.}~\bibnamefont{{van der Straten}}},
  \bibnamefont{and} \bibinfo{author}{\bibfnamefont{H.}~\bibnamefont{{Stoof}}},
  \bibinfo{journal}{ArXiv e-prints} \bibinfo{eid}{arXiv:1811.12835}
  (\bibinfo{year}{2018}), \eprint{1811.12835}.

\bibitem[{\citenamefont{{Hayata} and {Hidaka}}(2018)}]{Hayata2018}
\bibinfo{author}{\bibfnamefont{T.}~\bibnamefont{{Hayata}}} \bibnamefont{and}
  \bibinfo{author}{\bibfnamefont{Y.}~\bibnamefont{{Hidaka}}},
  \bibinfo{journal}{ArXiv e-prints}  (\bibinfo{year}{2018}),
  \eprint{1808.07636}.

\bibitem[{\citenamefont{{Surace} et~al.}(2018)\citenamefont{{Surace},
  {Russomanno}, {Dalmonte}, {Silva}, {Fazio}, and {Iemini}}}]{Surace2018}
\bibinfo{author}{\bibfnamefont{F.~M.} \bibnamefont{{Surace}}},
  \bibinfo{author}{\bibfnamefont{A.}~\bibnamefont{{Russomanno}}},
  \bibinfo{author}{\bibfnamefont{M.}~\bibnamefont{{Dalmonte}}},
  \bibinfo{author}{\bibfnamefont{A.}~\bibnamefont{{Silva}}},
  \bibinfo{author}{\bibfnamefont{R.}~\bibnamefont{{Fazio}}}, \bibnamefont{and}
  \bibinfo{author}{\bibfnamefont{F.}~\bibnamefont{{Iemini}}},
  \bibinfo{journal}{ArXiv e-prints} \bibinfo{eid}{arXiv:1811.12426}
  (\bibinfo{year}{2018}), \eprint{1811.12426}.

\bibitem[{\citenamefont{{Kreil} et~al.}(2018)\citenamefont{{Kreil},
  {Musiienko-Shmarova}, {Bozhko}, {Pomyalov}, {L'vov}, {Eggert}, {Serga}, and
  {Hillebrands}}}]{Kreil2018}
\bibinfo{author}{\bibfnamefont{A.~J.~E.} \bibnamefont{{Kreil}}},
  \bibinfo{author}{\bibfnamefont{H.~Y.} \bibnamefont{{Musiienko-Shmarova}}},
  \bibinfo{author}{\bibfnamefont{D.~A.} \bibnamefont{{Bozhko}}},
  \bibinfo{author}{\bibfnamefont{A.}~\bibnamefont{{Pomyalov}}},
  \bibinfo{author}{\bibfnamefont{V.~S.} \bibnamefont{{L'vov}}},
  \bibinfo{author}{\bibfnamefont{S.}~\bibnamefont{{Eggert}}},
  \bibinfo{author}{\bibfnamefont{A.~A.} \bibnamefont{{Serga}}},
  \bibnamefont{and}
  \bibinfo{author}{\bibfnamefont{B.}~\bibnamefont{{Hillebrands}}},
  \bibinfo{journal}{arXiv e-prints} \bibinfo{eid}{arXiv:1811.05801}
  (\bibinfo{year}{2018}), \eprint{1811.05801}.

\bibitem[{\citenamefont{Guo et~al.}(2013)\citenamefont{Guo, Marthaler, and
  Sch\"on}}]{Guo2013}
\bibinfo{author}{\bibfnamefont{L.}~\bibnamefont{Guo}},
  \bibinfo{author}{\bibfnamefont{M.}~\bibnamefont{Marthaler}},
  \bibnamefont{and} \bibinfo{author}{\bibfnamefont{G.}~\bibnamefont{Sch\"on}},
  \bibinfo{journal}{Phys. Rev. Lett.} \textbf{\bibinfo{volume}{111}},
  \bibinfo{pages}{205303} (\bibinfo{year}{2013}),
  \urlprefix\url{https://link.aps.org/doi/10.1103/PhysRevLett.111.205303}.

\bibitem[{\citenamefont{Guo and Marthaler}(2016)}]{Guo2016}
\bibinfo{author}{\bibfnamefont{L.}~\bibnamefont{Guo}} \bibnamefont{and}
  \bibinfo{author}{\bibfnamefont{M.}~\bibnamefont{Marthaler}},
  \bibinfo{journal}{New Journal of Physics} \textbf{\bibinfo{volume}{18}},
  \bibinfo{pages}{023006} (\bibinfo{year}{2016}),
  \urlprefix\url{http://stacks.iop.org/1367-2630/18/i=2/a=023006}.

\bibitem[{\citenamefont{Guo et~al.}(2016)\citenamefont{Guo, Liu, and
  Marthaler}}]{Guo2016a}
\bibinfo{author}{\bibfnamefont{L.}~\bibnamefont{Guo}},
  \bibinfo{author}{\bibfnamefont{M.}~\bibnamefont{Liu}}, \bibnamefont{and}
  \bibinfo{author}{\bibfnamefont{M.}~\bibnamefont{Marthaler}},
  \bibinfo{journal}{Phys. Rev. A} \textbf{\bibinfo{volume}{93}},
  \bibinfo{pages}{053616} (\bibinfo{year}{2016}),
  \urlprefix\url{https://link.aps.org/doi/10.1103/PhysRevA.93.053616}.

\bibitem[{\citenamefont{Pengfei et~al.}(2018)\citenamefont{Pengfei, Michael,
  and Guo}}]{Liang2017}
\bibinfo{author}{\bibfnamefont{L.}~\bibnamefont{Pengfei}},
  \bibinfo{author}{\bibfnamefont{M.}~\bibnamefont{Michael}}, \bibnamefont{and}
  \bibinfo{author}{\bibfnamefont{L.}~\bibnamefont{Guo}}, \bibinfo{journal}{New
  Journal of Physics} \textbf{\bibinfo{volume}{20}}, \bibinfo{pages}{023043}
  (\bibinfo{year}{2018}), ISSN \bibinfo{issn}{1367-2630},
  \urlprefix\url{http://stacks.iop.org/1367-2630/20/i=2/a=023043}.

\bibitem[{\citenamefont{{Lustig} et~al.}(2018)\citenamefont{{Lustig},
  {Sharabi}, and {Segev}}}]{Lustig2018}
\bibinfo{author}{\bibfnamefont{E.}~\bibnamefont{{Lustig}}},
  \bibinfo{author}{\bibfnamefont{Y.}~\bibnamefont{{Sharabi}}},
  \bibnamefont{and} \bibinfo{author}{\bibfnamefont{M.}~\bibnamefont{{Segev}}},
  \bibinfo{journal}{ArXiv e-prints}  (\bibinfo{year}{2018}),
  \eprint{1803.08731}.

\bibitem[{\citenamefont{Li et~al.}(2012)\citenamefont{Li, Gong, Yin, Quan, Yin,
  Zhang, Duan, and Zhang}}]{Li2012}
\bibinfo{author}{\bibfnamefont{T.}~\bibnamefont{Li}},
  \bibinfo{author}{\bibfnamefont{Z.-X.} \bibnamefont{Gong}},
  \bibinfo{author}{\bibfnamefont{Z.-Q.} \bibnamefont{Yin}},
  \bibinfo{author}{\bibfnamefont{H.~T.} \bibnamefont{Quan}},
  \bibinfo{author}{\bibfnamefont{X.}~\bibnamefont{Yin}},
  \bibinfo{author}{\bibfnamefont{P.}~\bibnamefont{Zhang}},
  \bibinfo{author}{\bibfnamefont{L.-M.} \bibnamefont{Duan}}, \bibnamefont{and}
  \bibinfo{author}{\bibfnamefont{X.}~\bibnamefont{Zhang}},
  \bibinfo{journal}{Phys. Rev. Lett.} \textbf{\bibinfo{volume}{109}},
  \bibinfo{pages}{163001} (\bibinfo{year}{2012}),
  \urlprefix\url{http://link.aps.org/doi/10.1103/PhysRevLett.109.163001}.

\bibitem[{\citenamefont{Huang et~al.}(2018{\natexlab{b}})\citenamefont{Huang,
  Li, and Yin}}]{Huang2017a}
\bibinfo{author}{\bibfnamefont{Y.}~\bibnamefont{Huang}},
  \bibinfo{author}{\bibfnamefont{T.}~\bibnamefont{Li}}, \bibnamefont{and}
  \bibinfo{author}{\bibfnamefont{Z.-q.} \bibnamefont{Yin}},
  \bibinfo{journal}{Phys. Rev. A} \textbf{\bibinfo{volume}{97}},
  \bibinfo{pages}{012115} (\bibinfo{year}{2018}{\natexlab{b}}),
  \urlprefix\url{https://link.aps.org/doi/10.1103/PhysRevA.97.012115}.

\bibitem[{\citenamefont{Lichtenberg and Lieberman}(1992)}]{Lichtenberg1992}
\bibinfo{author}{\bibfnamefont{A.}~\bibnamefont{Lichtenberg}} \bibnamefont{and}
  \bibinfo{author}{\bibfnamefont{M.}~\bibnamefont{Lieberman}},
  \emph{\bibinfo{title}{Regular and chaotic dynamics}}, Applied mathematical
  sciences (\bibinfo{publisher}{Springer-Verlag}, \bibinfo{year}{1992}), ISBN
  \bibinfo{isbn}{9783540977452},
  \urlprefix\url{https://books.google.pl/books?id=2ssPAQAAMAAJ}.

\bibitem[{\citenamefont{Buchleitner et~al.}(2002)\citenamefont{Buchleitner,
  Delande, and Zakrzewski}}]{Buchleitner2002}
\bibinfo{author}{\bibfnamefont{A.}~\bibnamefont{Buchleitner}},
  \bibinfo{author}{\bibfnamefont{D.}~\bibnamefont{Delande}}, \bibnamefont{and}
  \bibinfo{author}{\bibfnamefont{J.}~\bibnamefont{Zakrzewski}},
  \bibinfo{journal}{Physics reports} \textbf{\bibinfo{volume}{368}},
  \bibinfo{pages}{409} (\bibinfo{year}{2002}),
  \urlprefix\url{http://www.sciencedirect.com/science/article/pii/S0370157302002703}.

\bibitem[{\citenamefont{Steane et~al.}(1995)\citenamefont{Steane, Szriftgiser,
  Desbiolles, and Dalibard}}]{Steane95}
\bibinfo{author}{\bibfnamefont{A.}~\bibnamefont{Steane}},
  \bibinfo{author}{\bibfnamefont{P.}~\bibnamefont{Szriftgiser}},
  \bibinfo{author}{\bibfnamefont{P.}~\bibnamefont{Desbiolles}},
  \bibnamefont{and} \bibinfo{author}{\bibfnamefont{J.}~\bibnamefont{Dalibard}},
  \bibinfo{journal}{Phys. Rev. Lett.} \textbf{\bibinfo{volume}{74}},
  \bibinfo{pages}{4972} (\bibinfo{year}{1995}),
  \urlprefix\url{http://link.aps.org/doi/10.1103/PhysRevLett.74.4972}.

\bibitem[{\citenamefont{Roach et~al.}(1995)\citenamefont{Roach, Abele, Boshier,
  Grossman, Zetie, and Hinds}}]{Roach1995}
\bibinfo{author}{\bibfnamefont{T.~M.} \bibnamefont{Roach}},
  \bibinfo{author}{\bibfnamefont{H.}~\bibnamefont{Abele}},
  \bibinfo{author}{\bibfnamefont{M.~G.} \bibnamefont{Boshier}},
  \bibinfo{author}{\bibfnamefont{H.~L.} \bibnamefont{Grossman}},
  \bibinfo{author}{\bibfnamefont{K.~P.} \bibnamefont{Zetie}}, \bibnamefont{and}
  \bibinfo{author}{\bibfnamefont{E.~A.} \bibnamefont{Hinds}},
  \bibinfo{journal}{Phys. Rev. Lett.} \textbf{\bibinfo{volume}{75}},
  \bibinfo{pages}{629} (\bibinfo{year}{1995}),
  \urlprefix\url{https://link.aps.org/doi/10.1103/PhysRevLett.75.629}.

\bibitem[{\citenamefont{Sidorov et~al.}(1996)\citenamefont{Sidorov, McLean,
  Rowlands, Lau, Murphy, Walkiewicz, Opat, and Hannaford}}]{Sidorov1996}
\bibinfo{author}{\bibfnamefont{A.~I.} \bibnamefont{Sidorov}},
  \bibinfo{author}{\bibfnamefont{R.~J.} \bibnamefont{McLean}},
  \bibinfo{author}{\bibfnamefont{W.~J.} \bibnamefont{Rowlands}},
  \bibinfo{author}{\bibfnamefont{D.~C.} \bibnamefont{Lau}},
  \bibinfo{author}{\bibfnamefont{J.~E.} \bibnamefont{Murphy}},
  \bibinfo{author}{\bibfnamefont{M.}~\bibnamefont{Walkiewicz}},
  \bibinfo{author}{\bibfnamefont{G.~I.} \bibnamefont{Opat}}, \bibnamefont{and}
  \bibinfo{author}{\bibfnamefont{P.}~\bibnamefont{Hannaford}},
  \bibinfo{journal}{Quantum and Semiclassical Optics: Journal of the European
  Optical Society Part B} \textbf{\bibinfo{volume}{8}}, \bibinfo{pages}{713}
  (\bibinfo{year}{1996}),
  \urlprefix\url{http://stacks.iop.org/1355-5111/8/i=3/a=030}.

\bibitem[{\citenamefont{Westbrook et~al.}(1998)\citenamefont{Westbrook,
  Westbrook, Landragin, Labeyrie, Cognet, Savalli, Horvath, Aspect, Hendel,
  Moelmer et~al.}}]{Westbrook1998}
\bibinfo{author}{\bibfnamefont{N.}~\bibnamefont{Westbrook}},
  \bibinfo{author}{\bibfnamefont{C.~I.} \bibnamefont{Westbrook}},
  \bibinfo{author}{\bibfnamefont{A.}~\bibnamefont{Landragin}},
  \bibinfo{author}{\bibfnamefont{G.}~\bibnamefont{Labeyrie}},
  \bibinfo{author}{\bibfnamefont{L.}~\bibnamefont{Cognet}},
  \bibinfo{author}{\bibfnamefont{V.}~\bibnamefont{Savalli}},
  \bibinfo{author}{\bibfnamefont{G.}~\bibnamefont{Horvath}},
  \bibinfo{author}{\bibfnamefont{A.}~\bibnamefont{Aspect}},
  \bibinfo{author}{\bibfnamefont{C.}~\bibnamefont{Hendel}},
  \bibinfo{author}{\bibfnamefont{K.}~\bibnamefont{Moelmer}},
  \bibnamefont{et~al.}, \bibinfo{journal}{Phys. Scr. T}
  \textbf{\bibinfo{volume}{78}}, \bibinfo{pages}{7} (\bibinfo{year}{1998}).

\bibitem[{\citenamefont{Lau et~al.}(1999)\citenamefont{Lau, Sidorov, Opat,
  McLean, Rowlands, and Hannaford}}]{Lau1999}
\bibinfo{author}{\bibfnamefont{D.~C.} \bibnamefont{Lau}},
  \bibinfo{author}{\bibfnamefont{A.~I.} \bibnamefont{Sidorov}},
  \bibinfo{author}{\bibfnamefont{G.~I.} \bibnamefont{Opat}},
  \bibinfo{author}{\bibfnamefont{R.~J.} \bibnamefont{McLean}},
  \bibinfo{author}{\bibfnamefont{W.~J.} \bibnamefont{Rowlands}},
  \bibnamefont{and}
  \bibinfo{author}{\bibfnamefont{P.}~\bibnamefont{Hannaford}},
  \bibinfo{journal}{Eur. Phys. J. D} \textbf{\bibinfo{volume}{5}},
  \bibinfo{pages}{193} (\bibinfo{year}{1999}),
  \urlprefix\url{https://doi.org/10.1007/s100530050244}.

\bibitem[{\citenamefont{Bongs et~al.}(1999)\citenamefont{Bongs, Burger, Birkl,
  Sengstock, Ertmer, Rz\c{a}\.zewski, Sanpera, and Lewenstein}}]{Bongs1999}
\bibinfo{author}{\bibfnamefont{K.}~\bibnamefont{Bongs}},
  \bibinfo{author}{\bibfnamefont{S.}~\bibnamefont{Burger}},
  \bibinfo{author}{\bibfnamefont{G.}~\bibnamefont{Birkl}},
  \bibinfo{author}{\bibfnamefont{K.}~\bibnamefont{Sengstock}},
  \bibinfo{author}{\bibfnamefont{W.}~\bibnamefont{Ertmer}},
  \bibinfo{author}{\bibfnamefont{K.}~\bibnamefont{Rz\c{a}\.zewski}},
  \bibinfo{author}{\bibfnamefont{A.}~\bibnamefont{Sanpera}}, \bibnamefont{and}
  \bibinfo{author}{\bibfnamefont{M.}~\bibnamefont{Lewenstein}},
  \bibinfo{journal}{Phys. Rev. Lett.} \textbf{\bibinfo{volume}{83}},
  \bibinfo{pages}{3577} (\bibinfo{year}{1999}),
  \urlprefix\url{https://link.aps.org/doi/10.1103/PhysRevLett.83.3577}.

\bibitem[{\citenamefont{Sidorov et~al.}(2002)\citenamefont{Sidorov, McLean,
  Scharnberg, Gough, Davis, Sexton, Opat, and Hannaford}}]{Sidorov2002}
\bibinfo{author}{\bibfnamefont{A.}~\bibnamefont{Sidorov}},
  \bibinfo{author}{\bibfnamefont{R.}~\bibnamefont{McLean}},
  \bibinfo{author}{\bibfnamefont{F.}~\bibnamefont{Scharnberg}},
  \bibinfo{author}{\bibfnamefont{D.}~\bibnamefont{Gough}},
  \bibinfo{author}{\bibfnamefont{T.}~\bibnamefont{Davis}},
  \bibinfo{author}{\bibfnamefont{B.}~\bibnamefont{Sexton}},
  \bibinfo{author}{\bibfnamefont{G.}~\bibnamefont{Opat}}, \bibnamefont{and}
  \bibinfo{author}{\bibfnamefont{P.}~\bibnamefont{Hannaford}},
  \bibinfo{journal}{Acta Phys. Pol. B} \textbf{\bibinfo{volume}{33}},
  \bibinfo{pages}{2137} (\bibinfo{year}{2002}).

\bibitem[{\citenamefont{Fiutowski et~al.}(2013)\citenamefont{Fiutowski,
  Bartoszek-Bober, Dohnalik, and Kawalec}}]{Fiutowski2013}
\bibinfo{author}{\bibfnamefont{J.}~\bibnamefont{Fiutowski}},
  \bibinfo{author}{\bibfnamefont{D.}~\bibnamefont{Bartoszek-Bober}},
  \bibinfo{author}{\bibfnamefont{T.}~\bibnamefont{Dohnalik}}, \bibnamefont{and}
  \bibinfo{author}{\bibfnamefont{T.}~\bibnamefont{Kawalec}},
  \bibinfo{journal}{Opt. Commun.} \textbf{\bibinfo{volume}{297}},
  \bibinfo{pages}{59} (\bibinfo{year}{2013}).

\bibitem[{\citenamefont{Kawalec et~al.}(2014)\citenamefont{Kawalec,
  Bartoszek-Bober, Pana\'{s}, Fiutowski, P{\l}awecka, and
  Rubahn}}]{Kawalec2014}
\bibinfo{author}{\bibfnamefont{T.}~\bibnamefont{Kawalec}},
  \bibinfo{author}{\bibfnamefont{D.}~\bibnamefont{Bartoszek-Bober}},
  \bibinfo{author}{\bibfnamefont{R.}~\bibnamefont{Pana\'{s}}},
  \bibinfo{author}{\bibfnamefont{J.}~\bibnamefont{Fiutowski}},
  \bibinfo{author}{\bibfnamefont{A.}~\bibnamefont{P{\l}awecka}},
  \bibnamefont{and} \bibinfo{author}{\bibfnamefont{H.-G.}
  \bibnamefont{Rubahn}}, \bibinfo{journal}{Opt. Lett.}
  \textbf{\bibinfo{volume}{39}}, \bibinfo{pages}{2932} (\bibinfo{year}{2014}),
  \urlprefix\url{http://ol.osa.org/abstract.cfm?URI=ol-39-10-2932}.

\bibitem[{\citenamefont{{Dutta, O. and Gajda, M. and Hauke, P. and Lewenstein,
  M. and L\"uhmann, D.-S. and Malomed, B. A. and Sowi\'nski, T. and Zakrzewski,
  J.}}(2015)}]{Dutta2015}
\bibinfo{author}{\bibnamefont{{Dutta, O. and Gajda, M. and Hauke, P. and
  Lewenstein, M. and L\"uhmann, D.-S. and Malomed, B. A. and Sowi\'nski, T. and
  Zakrzewski, J.}}}, \bibinfo{journal}{Reports on Progress in Physics}
  \textbf{\bibinfo{volume}{78}}, \bibinfo{pages}{066001}
  (\bibinfo{year}{2015}), ISSN \bibinfo{issn}{0034-4885},
  \urlprefix\url{http://stacks.iop.org/0034-4885/78/i=6/a=066001}.

\bibitem[{\citenamefont{Pethick and Smith}(2002)}]{Pethick2002}
\bibinfo{author}{\bibfnamefont{C.}~\bibnamefont{Pethick}} \bibnamefont{and}
  \bibinfo{author}{\bibfnamefont{H.}~\bibnamefont{Smith}},
  \emph{\bibinfo{title}{{Bose-Eistein condensation in dilute gases}}}
  (\bibinfo{publisher}{{Cambridge University Press}},
  \bibinfo{address}{{Cambridge, England}}, \bibinfo{year}{2002}).

\bibitem[{\citenamefont{{{\"O}hberg} and {Wright}}(2018)}]{Ohberg2018}
\bibinfo{author}{\bibfnamefont{P.}~\bibnamefont{{{\"O}hberg}}}
  \bibnamefont{and} \bibinfo{author}{\bibfnamefont{E.~M.}
  \bibnamefont{{Wright}}}, \bibinfo{journal}{arXiv e-prints}
  \bibinfo{eid}{arXiv:1812.04672} (\bibinfo{year}{2018}), \eprint{1812.04672}.

\end{thebibliography}

\end{document}